\documentclass[pra,twocolumn,showpacs,aps,superscriptaddress]{revtex4-1}
 
\usepackage{amsmath}
\usepackage{amssymb}
\usepackage{amsmath}
\usepackage{txfonts}
\usepackage{graphicx}
\usepackage{epsfig}
\usepackage[T1]{fontenc}
\usepackage{times}
\usepackage{color} 
\usepackage{xspace}
\usepackage{amssymb,amsmath}
\usepackage{amsbsy}
\usepackage{braket}
\usepackage{tabularx}
\usepackage{mathtools}        %loads amsmath as well

\usepackage{ifpdf}
\ifpdf
\usepackage{epstopdf}
\fi

\usepackage[unicode]{hyperref}
\hypersetup{
	pdftitle={vortex},
	pdfauthor={Chenjiawen},
	pdfsubject={vortex},
	colorlinks=true,
	linkcolor=blue,
	citecolor=blue,
	filecolor=black,
	urlcolor =blue
}
\def\urlprefix{}
\def\url#1{}

\usepackage{bm}% bold math
\usepackage{color}
\newcommand{\be}{\begin{equation}}

\newcommand{\ee}{\end{equation}}

\newcommand{\bea}{\begin{eqnarray}}

\newcommand{\eea}{\end{eqnarray}}

\renewcommand{\d}{{\rm d }}

\newcommand{\p}{\partial}

\newcommand{\nn}{\nonumber }

\newcommand{\st}{\theta}

\newcommand{\dt}{\delta}

\newcommand{\s}{{\sigma}}

\newcommand\smallO{
	\mathchoice
	{{\scriptstyle\mathcal{O}}}% \displaystyle
	{{\scriptstyle\mathcal{O}}}% \textstyle
	{{\scriptscriptstyle\mathcal{O}}}% \scriptstyle
	{\scalebox{.7}{$\scriptscriptstyle\mathcal{O}$}}%\scriptscriptstyle
}

\begin{document}

%\title{Statistical mechanics of vortices clusters on a sphere } 
\title{Onsager vortex clusters on a sphere } 

\author{Jiawen Chen}
\affiliation{Graduate School of China Academy of Engineering Physics, Beijing 100193, China}
\author{Xiaoquan Yu}
\email{xqyu@gscaep.ac.cn}

\affiliation{Graduate School of China Academy of Engineering Physics, Beijing 100193, China}
\affiliation{Department of Physics, and Centre for Quantum Science, University of Otago, Dunedin 9016, New Zealand}

%\affiliation{ }
%%

\begin{abstract}
We study Onsager vortex clustered states in a shell-shaped superfluid containing a large number of  quantum vortices.  In the  incompressible limit and at low temperatures,  the relevant problem can be  boiled down to the statistical mechanics of neutral point vortices confined on a sphere.  We analyze  rotation free vortex clustered states within the mean field theory in the microcanonical ensemble. We find that the sandwich state, which involves the separating of vortices with opposite circulation and the clustering of vortices with the same circulation around the poles and the equator, is the maximum entropy vortex distribution, subject to zero angular momentum constraint. 
The dipole momentum vanishes for the sandwich state and the quadrupole tensor serves as an order parameter to characterize the vortex cluster structure.  
For given finite angular momentum,  the equilibrium vortex distribution forms a dipole structure, i.e.,  vortices with opposite sign are separated and are accumulated around the south and north pole, respectively.
The conditions for the onset of clustering, and the exponents associated with the quadrupole moment and the dipole moment as functions of energy, are obtained within the mean field theory.  At large energies,  
 we obtain asymptotically exact vortex density distributions using the stereographic projection method,  giving rise to the parameter bounds for the vortex clustered states. 
The analytical predictions are in excellent agreement with microcanonical Monte Carlo simulations.

 \end{abstract}

\maketitle

\section{Introduction}
Coherent large vortex structures can occur in various bounded flows,  examples range from Great Red Spot in Jupiter's atmosphere~\cite{GreatRedSpot}, to giant vortex clusters in atomic Bose-Einstein condensates (BECs)~\cite{Gauthier2019, Johnstone2019}, and in dissipative quantum fluids of exciton-polaritons~\cite{panico2023onset}. 
For a vortex system in a bounded domain, the phase space per vortex is also bounded.  The formation of vortex clusters is due to that the system favors the clustering of like-sign vortices at high energies \cite{1949Onsager,EyinkRMP}.  Onsager formulates the statistical mechanics of a bounded point vortex system in the microscopic ensemble and demonstrates that the clustered state is at negative absolute temperature.  Since then clustering of vortices on bounded domains has been investigated extensively~\cite{Joyce1973, edwards1974negative, williamson1977, Pointin, campbell1991statistics, SmithPRL, SmithONeil, Kiessling1995, Yatsuyanagi2005, Esler2015, Qi2014, siggia1981,  Matt2012, TomPRA, TapioPRL, Gurarie2004, Kiessling2012,  clusteringYu,  HayderPRA2016, Valani2018, Junsik2018, Junsik2019, AnghelutaPRE,  AudunPRE2017, Guo2021, Patil2021, review2017, chavanis1996, Chavanis2002, Xiong2023iop}.  The clustering of vortices is also closely related to the end state of inverse energy cascades~\cite{DecayingQTBillam, AnghelutaPRE,IECMatt}, which involves energy transport from small to large scales in two-dimensional(2D) turbulence. 
 
Previous studies on Onsager vortices  mainly focus on systems confined on a flat region. Recent experimental realizations of  BECs~\cite{aveline2020observation}  and ultracold atomic bubbles~\cite{carollo2022observation} in International Space Station open a possibility to investigate a bubble trapped superfluid and vortex physics  on curved surfaces experimentally. Curvature and topology of a surface have non-trivial effects on BECs~\cite{Tononi2019, Tononi2020, tononi2023shellshaped}, vortex dynamics~\cite{Bogomolov1978, stuart2004, Hally, polvani1993, Kimura99, Newton2001TheNP, Alobaidi2006, Hassan2007, Vitelli2004, Turner2010, dritschel2015motion,Drit2015, Doshi2021, PhysRevA.103.053306, PhysRevA.105.023307, PhysRevA.102.043305,Saito2023, GuangyaoPRA, Xiong2024}, and the Berezinskii-Kosterlitz-Thouless (BKT) transition~\cite{PhysRevResearch.4.013122}.  The formation and structure of Onsager vortices also strongly depend on the surface geometry.  On a sphere, 
 the distributions of single species of vortices in the background of uniformly distributed vortices with the opposite circulation have been studied in details~\cite{Kiessling2012}.  Recent numerical simulations of the point vortex model with removing low energy vortex-antivortex pairs show that, different from the giant dipole configuration for vortices confined to a disc~\cite{TapioPRL,clusteringYu}, in a spherical shell trap uniformly distributed  neutral vortices form a quadruple cluster~\cite{Guo2021}.  However, it is not clear whether the quadruple clustered state is the statistically most favored state for given energy, angular momentum and  vortex number.  To answer this question requires the study of distributions of two species of vortices on a sphere simultaneously, which has not been well-addressed. 

In this paper, we systematically investigate rotation-free clustered states of  a neutral vortex system confined on a sphere.  We extend the mean field theory of vortex statistical mechanics developed by Joyce and Montgomery~\cite{Joyce1973} to 
a sphere and treat the distributions of the two species of vortices simultaneously.   In particular,  we analyze the self-consistent equation of the stream function on a sphere near the uniform state and in the high energy limit.
We find that:  under the constraint of zero angular momentum, the sandwich state where vortices with opposite sign are distributed around the poles and the equator respectively, is the maximum entropy state and has higher statistical weight than the quadrupole state; for finite angular momentum,  a giant dipole structure is statistically favored.   The sandwich state and the dipole state are characterized by the non-vanishing quadrupole moment and the dipole moment, respectively.  The energy dependence and the upper bounds of the quadrupole moment and the dipole moment at low energies and in the high energy limit are also obtained within the mean-field theory.  To test the mean-field predictions, we perform the microcanonical Monte Carlo (MC) sampling of finite vortex numbers and find good agreement on the energy dependence of the quadrupole moment and the dipole moment for a system containing a total of $N = 1000$ vortices and the values of the moments showing  a saturation to the predicted upper bounds.

\section{Point vortices on a sphere} 
For a superfluid on a plane, the circulation of a quanutm vortex is quantized  in units of circulation quantum $\kappa \equiv2 \pi \hbar/m$~\cite{BECbook}, and the vorticity has a singularity at the vortex core $\bm{r}_i$: $\omega(\bm{r})=\nabla \times \bm{u}=\kappa_i \delta(\bm{r}-\bm{r}_i)=\kappa\sigma_i \delta(\bm{r}-\bm{r}_i)$  with sign $\sigma_i=\pm 1$ for singly charged vortices.  Here $m$ is the atomic mass and $\bm{u}$ is the superfluid velocity. The quantization ensures that the vorticity of a quantum vortex concentrates around the core region~\cite{EyinkRMP}.  Hence when the mean separation  between quantum vortices $\bar{d}$  is much larger than the vortex  core size $\xi$~\cite{Fetter1998,Barenghi_2016}, the point vortex model (PVM) describes well the dynamics of quantum vortices in the incompressible limit~\cite{1949Onsager,EyinkRMP,novikov1975,Aref1999},  provided vortex annihilation can be neglected.  In this limit, the flow is nearly incompressible $\nabla \cdot \bm{u}=0$ and hence a stream function $\psi$ can be introduced such that $\bm{u}=\nabla \psi \times \hat{\bm{z}}$, where $\hat{\bm{z}}$ is the unit vector normal to the plane.  The stream function is completely determined by  vortex positions and $\psi(\bm{r})=-\sum_{i}\kappa_i G(\bm{r},\bm{r}_i)$, where $G(\bm{r},\bm{r}_i)=-(1/2\pi)\log(|\bm{r}-\bm{r}_i|/\xi)$ is the Green's function of Laplacian operator in 2D:  $\nabla^2 G(\bm{r},\bm{r}_i)=-\delta(\bm{r}-\bm{r}_i)$.    
The PVM also describes 2D superfluid transition at finite temperature, which is the celebrated  BKT transition \cite{Kosterlitz_1973, Kosterlitz_1974}.

On curved surfaces, a covariant PVM has been well formulated~\cite{Hally, Kimura99, dritschel2015motion}. Here we only focus on the PVM on a sphere.  The Riemannian  metric on a sphere is 
\bea
{\rm d} s^2=R^2 {\rm d}\theta^2+R^2\sin^2\theta {\rm d} \phi^2, 
\eea 
where $R$ is the radius,  $\theta$ is the polar angle and  $\phi$ is the azimuthal angle.

\begin{figure}[htp]
	\centering
	\includegraphics[width=6cm]{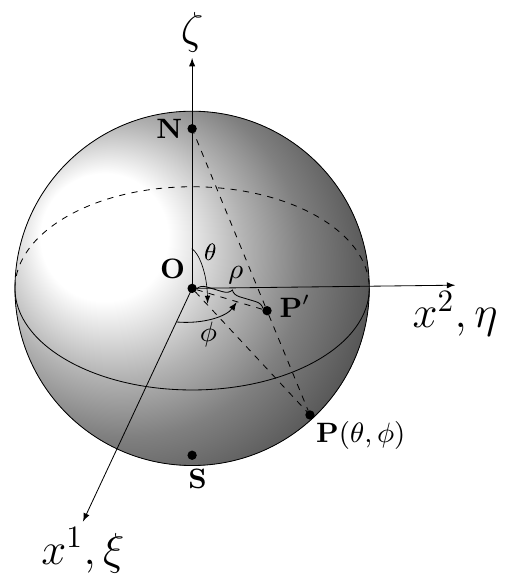}
	\caption{Stereographic projection from the north pole.  The point ${\rm P}(\st,\phi)$ on the southern hemisphere is projected onto the $\zeta=0$ plane and is denoted as $ {\rm P}^\prime$. }
	\label{proj}
\end{figure}

The Cartesian coordinates relate to the spherical coordinates via
\bea
\xi=R \sin \theta \cos \phi, \quad  \eta=R \sin \theta \sin \phi, \quad \zeta=R \cos \theta.
\eea
The projection from the north pole to the $\zeta=0$ plane induces stereographic coordinates $z=x^{1}+ {\rm i}x^{2}$, where $x^1$ and $x^2$ axes are identified as the $\xi$ and $\eta$ axes, respectively (Fig.~\ref{proj}). Let us also introduce the following notations    
$\partial_z \equiv (\partial_1-{\rm i}\partial_2)/2$ and $\partial_{\bar{z}}\equiv(\partial_1+ {\rm i}\partial_2)/2$ . 
In terms of stereographic coordinates, the metric reads
${\d} s^2=h(x^1,x^2)[(\d x^{1})^2+(\d x^{2})^2]$, where $h=4 R^4/(R^2+|z|^2)^2$.
For a sphere, stereographic coordinates are isothermal coordinates~\cite{chern1955elementary} and are related to the spherical coordinates by
\bea
z=R \cot \left(\frac{\theta}{2}\right) {\rm e}^{\rm i \phi }.
\eea
It is sometimes convenient to introduce polar coordinates $z=\rho {\rm e}^{ \rm i \phi}$, where $\rho=R\cot (\theta/2)$ and the polar angle $\phi$ is chosen to coincide with  the azimuthal angle.
The relations between the three coordinate systems are 
\bea
\xi&=&\frac{2x^{1} R^2}{R^2+|z|^2}=\sqrt{h}x^{1}=R \sin \theta \cos \phi, \\
\eta&=&\frac{2 x^{2} R^2}{R^2+|z|^2}=\sqrt{h}x^{2}=R\sin \theta \sin \phi ,\\
\zeta&=&R\frac{R^2-|z|^2}{R^2+|z|^2}= R \cos \theta,
\eea
and we use Cartesian, spherical or stereographic coordinates wherever convenient. 

In spherical coordinates, the stream function reads 
\bea 
\psi(\theta,\phi)=\kappa\sum_{i} \sigma_i G(\theta,\phi; \theta_i,\phi_i),
\label{streamfunction}
\eea
where $(\theta_i,\phi_i)$ is the position of the $i$th vortex, $G(\theta,\phi; \theta_i,\phi_i)$ is the Green's function satisfying~\cite{Kimura99, dritschel2015motion}
\bea 
\nabla^2 G(\theta,\phi; \theta_i,\phi_i)=-\delta_s(\theta,\phi; \theta_i,\phi_i)+1/\Omega,
\label{Greenfunction}
\eea
where
\bea
\delta_s(\theta,\phi; \theta_i,\phi_i)=\frac{1}{R^2 \sin \theta}\delta(\theta-\theta_i)\delta(\phi-\phi_i),
\eea
$\Omega=4\pi R^2$ is the area of the sphere,  and 
$\nabla^2=(1/\sin \theta)\partial_{\theta}(\sin \theta \partial_\theta)+( 1/\sin^2 \st )\partial^2_{\phi}$ is the Laplace-Bertrami operator.
The Green's function reads~\cite{Kimura99}
\bea
G(\theta, \phi; \theta_i,\phi_i)&&=-\frac{1}{2\pi} \log \frac{\sqrt{2R^2- 2\bm{r} \cdot \bm{r}_i } }{\xi} \nn\\
&&=-\frac{1}{2\pi} \log \left(\frac{\sqrt{2}R}{\xi}\sqrt{1-\cos \Theta }\right), 
\eea
where $\Theta$ is the angle between $\bm{r}$ and $\bm{r}_i$, and $\cos \Theta=\cos \theta \cos \theta_i + \sin \theta \sin \theta_i \cos (\phi-\phi_i)$. It is interesting to notice  that $\sqrt{2R^2-2\bm{r} \cdot \bm{r}_i}$ is the Euclidean distance, but not the geodesic distance, between two points on the sphere.

On a closed surface, the total vorticity must vanish, namely,  $\sum_i \sigma_i=0$.  Hence combining Eqs.~\eqref{streamfunction} and \eqref{Greenfunction}, we obtain
\bea 
\nabla^2 \psi(\theta,\phi)=-\omega(\theta,\phi),
\eea
where 
\bea
\omega(\theta,\phi)=\kappa \sum_{i}  \sigma_i \delta(\theta,\phi; \theta_i,\phi_i).
\label{vorticity}
\eea
The Hamiltonian of the point vortices on a sphere is 
\bea
H=\frac{\rho_s \kappa^2}{2}\sum_{i\neq j} \sigma_i \sigma_j G(\theta_i, \phi_i; \theta_j,\phi_j). 
\label{Hamiltonian}
\eea
On a sphere, besides the energy and the number of vortices, the following quantities 
\bea
L_{\xi}&&=\kappa R^2 \sum_i \sigma_i \sin\theta_i \cos\phi_i , \\
L_{\eta}&&=\kappa R^2 \sum_i \sigma_i\sin\theta_i \sin\phi_i,  \\
L_{\zeta}&&=\kappa R^2 \sum_i  \sigma_i \cos\theta_i,
\eea 
are also conserved and are associated with the fluid angular momentum.

\section{Onsager vortex clusters on a sphere}

\subsection{Onsager vortex clusters at negative temperature}
A particular  feature of 2D point-vortex systems in a bounded domain is that the phase space area per vortex  is finite.  For a vortex system on a  sphere,  the phase space of a vortex is  the spherical surface. Let us subdivide the surface  into  small cells of area  $a_i \ll \Omega$, where $\Omega$ is the surface area.  While the cells are still  large enough to  contain many vortices.
We consider a system consisting of $N_{+}$ positive vortices and $N_{-}=N_{+}$ negative vortices. The total vortex number is $N=N_++N_-$. We denote the number of vortices in a cell of each species as  $N_{i,\hat{\sigma}}$ with $\hat{\sigma}=\pm$ and the number of states is then given by~\cite{Joyce1973, Maestrini2019}
\begin{align}
	\mathcal{W}= \prod_{\hat{\sigma}} \left\{ \prod_i \frac{N_{\hat{\sigma}}!}{N_{i,\hat{\sigma}}!}\left(\frac{a_i}{\Omega} \right)^{N_{i,\hat{\sigma}}} \right\} .
	\label{numberofstates}
\end{align}
Then the thermodynamic entropy is $k_B\log \mathcal{W}(H)$ , where $k_B$ is the Boltzmann constant.  Since $k_B\log \mathcal{W}(-\infty)=k_B\log \mathcal{W}(+\infty)=0$, which correspond to extreme situations of  vortex-antivortex pairs  collapsing  and  like-sign vortices concentrating  into a point, respectively.  Then $\mathcal{W}$ must reach its maximum at an intermediate energy $H_m$ and the inverse temperature $1/T\equiv k_B\partial \log \mathcal{W}(H)/\partial H$
is thus negative for $H>H_m$. High energy states at negative temperature correspond to the formation of macroscopic vortex  clusters~\cite{1949Onsager,EyinkRMP}. 

For given total vortex numbers $N_{\pm}$ and the area $\Omega$, the entropy is a function of $\{ N_{i,\pm}\}$ and for large $N_{i,\pm}$  we obtain 
\bea
(1/N_{\pm})\log \mathcal{W} &&\simeq  -\sum_{i} \left(\frac{N_{i,+}}{N_{+}a_{i}} \log \frac{N_{i,+}}{N_{+}a_{i}} \right)a_i\notag\\ 
&&-\sum_{i} \left(\frac{N_{i,-}}{N_{-}a_{i}} \log \frac{N_{i,-}}{N_{-}a_{i}} \right)a_i,
\label{statenumber}
\eea
where the Stirling's formula is applied and the irrelevant constants are ignored.

\subsection{Continuous limit and the mean-field theory}
The PVM Eq.~\eqref{Hamiltonian} exhibits Onsager clustered states at high energies.  In clustered phases, the energy $H \sim N^2$, hence in order to describe clustered phases in the proper thermodynamics limit,  we need to make the rescaling $\sigma_i \rightarrow \sigma_i/N_{\pm}$ and define 
the vortex number densities on a sphere as following:
\begin{align}
 n_{\pm }(\bm{r})&=\frac{1}{N_{\pm}}\sum^{ N_{\pm}}_{i } \delta_s(\theta,\phi; \theta_i,\phi_i),  
\end{align}  
where $n_{\pm}$ satisfy the normalization condition $\int \d \Omega  \, n_\pm =1$.
Then the rescaled vortex charge density (vorticity) is 
\bea
\omega(\bm{r})=\kappa\s(\bm{r})\equiv\kappa [n_+(\bm{r})-n_-(\bm{r})],
\eea 
and the rescaled stream function 
\bea
 \psi(\bm{r})=\kappa \int \d\Omega' G(\bm{r},\bm{r}')\,\sigma(\bm{r}'), 
\eea
satisfying 
\bea
\nabla^2 \psi(\bm{r}) = -\kappa \s (\bm{r}).  
\label{n1}
\eea
In the following we consider vortex distributions at large scales and treat $\sigma$, $n_{\pm}$ and $\psi$ as smooth functions at scales much larger than the mean seperation $\bar{d}$.  The rescaled  coarse-grained Hamiltonian reads
\bea
 E&&=\frac{\rho_s \kappa^2}{2}\int \d \Omega \d \Omega' \sigma(\bm{r})G(\bm{r},\bm{r}')\sigma(\bm{r}') \nn\\
&&=\frac{\rho_s}{2}\int \d\Omega  \, \omega(\bm{r})\psi (\bm{r}), 
\label{energy}
\eea 
and $E\sim \mathcal{O}(1)$.
For a sphere, this Hamiltonian is nothing but the rescaled kinetic energy of the flow generated by vortices: $E=\rho_s/2\int \d\Omega  \, |\bm{u}|^2=\rho_s/2\int \d\Omega  \, |\nabla \psi|^2=\rho_s/2\int \d\Omega \, \omega \, \psi$, 
where the fluid velocity $\bm u=\nabla \psi \times \bm{e}_r $ and $\bm e_r$ is the radial unit vector.  Applying point vortices directly to Eq.~\eqref{energy} [plugging Eqs.~\eqref{streamfunction} and \eqref{vorticity} into Eq.~\eqref{energy}], we obtain that $E= H/N^2_{\pm }+H_{\text{singular}}$, where $H_{\text{singular}}=(\rho_s\kappa^2/4\pi N^2_{\pm}) \sum_i  \sigma^2_i \log [d(\bm{r}_i,\bm{r}_i)/\xi]$, and $d(\bm{r}_i,\bm{r}_j)$ is the Euclidean distance between the two points on a sphere.  For point vortices, $H_{\text{singular}}=\infty$.  In practice,  the distance between vortices can not be smaller than the vortex core size $\xi$ which serves the ultraviolet cut-off of the PVM. By choosing $d(\bm{r}_i,\bm{r}_i)=\xi$, we have $H_{\text{singular}}=0$. 

In terms of the rescaled collective variables of vortices, the components of the angular momentum read 
\bea
L_{\xi}&&=\kappa R^2 \int \d \Omega\, \sigma \sin\theta \cos\phi , \\
L_{\eta}&&= \kappa R^2 \int \d \Omega\, \sigma \sin\theta \sin\phi,  \\
L_{\zeta}&&= \kappa R^2 \int \d  \Omega \, \sigma \cos\theta.
\eea  
Hereafter we choose the $\zeta$-axis in the direction of total angular momentum and hence $ \bm{L}=L_{\zeta}$ and  we set $\rho_s=\kappa=R=k_B=1$ for convenience. In this convention $\xi \ll \bar{d} \ll 1$.

In the continuous limit, $a_i\rightarrow 0$ and $N_{i,\pm}/N_{\pm}a_{i} \rightarrow n_{\pm}(\bm{r})$, and the Eq.~\eqref{statenumber} gives rise to  the thermodynamic entropy per vortex  for given vortex distributions $n_{\pm}$~\cite{Joyce1973, EyinkRMP}: 
\begin{align}
 S[n_+,n_-]=- \int  \d \Omega \left(n_+\log n_++n_-\log n_- \right),
 \label{entropy}
 \end{align}  
 and the most probable vortex density distribution is given by maximizing the entropy Eq.~\eqref{entropy} subject to given energy $E$, vortex number $N_{\pm}$ and angular momentum $L_{\zeta}$:
\begin{align}
	\delta S-\beta \delta E-\alpha \delta L_{\zeta}-\mu_+ \delta N_+/N_+ -\mu_- \delta N_-/N_-=0,
	 \label{maxentropy}
\end{align}
where $\beta$ and $\mu_\pm$ are Lagrange multipliers and have the interpretation of inverse temperature and chemical potentials in microcanonical ensemble, respectively.  
Here $\alpha$ is the Lagrange multiplier associated with the angular momentum and $\omega\equiv\alpha /\beta$ has the meaning of  rotation frequency.

The variation equation~\eqref{maxentropy} gives rise to
\begin{align}
	 n_\pm=\exp \left[ \mp \beta \psi(\bm r) \mp \alpha \cos \st+\gamma_\pm \right].  
	\label{n2}   
\end{align} 
 where$\gamma_\pm =-\mu_\pm-1$. Combining Eq.~\eqref{n1}
 we obtain the self-consistent equation for determining the structure of coherent Onsager vortex clusters:  
 \bea
  \label{selfconsistent}
 \nabla^2 \psi(\bm{r})&=&\exp[\beta \psi(\bm{r} )+\alpha \cos \st+\gamma_-]\\
 &&-\exp[ - \beta \psi(\bm r)-\alpha \cos \st+\gamma_+].\nn
\eea   
Non-uniform solutions at negative temperatures to Eq.~\eqref{selfconsistent} describe Onsager vortex states.  

\subsection{Onset  of  clustering}
\label{clustering}

The onset of clustering involves global eigen-modes of the Laplace-Bertrami operator.  It is then convenient to use spherical coordinates.  
Let us consider a solution to Eq.~\eqref{selfconsistent} for the given values of $E$ and  $L_{\zeta}$, and a nearby solution $n_\pm +\dt n_\pm$ at $E +\dt E$ and $L_{\zeta}+\dt L_{\zeta}$. The corresponding changes in the constraints are
\begin{align}
 		\label{c1}
 	0&=\int \d \Omega \ \dt n_\pm ,\\ 
 		\label{c2}
 \delta E &=\int \d\Omega\  \psi \delta \sigma  +\frac12 \int \d \Omega \, \dt \s  \dt \psi, \\
 	\label{c3}
 	\delta L_{\zeta} &=\int \d \Omega\  \delta \sigma \cos \st  \ .
 \end{align}  
For the homogeneous state $n_\pm=n_0=1/4\pi$, and $\alpha=\sigma=\psi=0$. Then Eqs.~\eqref{c1},\eqref{c2} and \eqref{c3} become 
\begin{align}
	\label{c11}
	{\rm e}^{-\delta \gamma_{\pm}}&=n_0\int \d \Omega \, \exp(\mp \beta \delta \psi+\delta \alpha \cos \theta),\\ 
	\label{c22}
	\delta E &=\frac{1}{2} \int \d \Omega \, \dt \s  \dt \psi, \\
	\label{c33}
	\delta L_{\zeta} &=\int \d \Omega\  \delta \sigma \cos \st  \ .
\end{align}  
Expanding $\delta \sigma$ and Eq. \eqref{c1} up to the leading order in $\delta \psi$, we have 
 \begin{align}
 	\label{c0}
 	 \delta \sigma&= -2n_0(\beta \dt \psi-\dt \gamma_+ + \cos \st  \delta \alpha), \\
 \delta \gamma_+&=- \delta \gamma_- =\beta n_0 \int \d \Omega \, \delta \psi,  
 	\label{c222}
 \end{align} 
and as it follows  Eqs.~\eqref{c22} and ~\eqref{c33} become
\begin{align} 	
 	\delta E &=-n_0 \int \d \Omega \, (\beta \dt \psi-\dt \gamma_+ + \cos \st  \delta \alpha) \dt \psi, \\
 	\label{c333}
 	 \dt L_{\zeta}&= -2n_0 \beta \int \d\Omega  \cos \st \, \dt \psi -\frac{2}{3}\dt \alpha,
 \end{align}  
where $\dt \alpha$ is assumed to be the same order as $\dt\psi$.
 
Combining Eqs.~\eqref{n1} and~\eqref{c0}, up to  leading order in $\delta \psi$, we obtain 
\bea
  \label{vLaplace}
  \nabla^2 \dt \psi &&=-\dt \s(\bm r) =2n_0(\beta \dt \psi-\dt \gamma_+ + \cos \st  \delta \alpha).  
\eea  
%   \dt E =& \int \Big[-n\beta \psi\dt \psi-n\psi^2 \dt  \beta+n_+\psi \dt \gamma_+ \notag \\ 
 
%We consider vortex clustered states near the homogeneous state $n_\pm=n_0=1/4\pi$. 
%For the homogeneous state $\alpha=0$ and $\sigma=\psi=0$, then  Eqs.~\eqref{c1},\eqref{c2} and \eqref{c3} become 
%\begin{align}
%    0=& \dt \gamma_++\dt \gamma_-,\\
%     0= &\dt \gamma_+-\dt \gamma_- -2\beta n_0 \int d \Omega \, \delta \psi  ,\\ 
%      \dt E=& \frac12 \int d \Omega \, \dt \s  \, \dt \psi,\\
%     \dt L_{\zeta}=& -2n_0 \beta \int d\Omega  \cos \st \, \dt \psi -\frac{2}{3}\dt \alpha,
%\end{align}
%where $\dt \alpha$ is assumed to be the same order as $\dt\psi$.
Let us express the changes of the Lagrange multipliers in terms of the changes of the corresponding conserved quantities and introduce the operator $\mathcal{P}$:
\begin{align}
\mathcal{P}\dt \psi &\equiv  \nabla^2 \dt \psi-2n_0 \bigg[\beta \dt \psi-n_0 \beta \int  \d \Omega  \, \dt \psi    \\
    &-3n_0 \beta \cos \st\int \d \Omega \cos \st \dt\psi -\frac{3}{2}  \cos \st \dt L_{\zeta}\notag\bigg].
\end{align}  
Then Eq.~\eqref{vLaplace} becomes a zero mode equation of  the  operator $\mathcal{P}$, i.e., 
\bea
\mathcal{P} \delta \psi=0.
\label{zeromode}
\eea

In order to solve the zero mode equation, we consider an expansion
 \begin{align} 
\dt \psi &=\sum^{\infty}_{\ell=0}\sum^{\ell}_{m=-\ell} \epsilon f_{\ell m}\psi_{\ell m}(\st,\phi),
\end{align}
where $\psi_{\ell m}$ satisfies $\mathcal P \psi_{\ell m}=0$, $f_{\ell m}$ is the mode coefficient, and $\epsilon \ll 1$ is a small amplitude. 
We denote $\delta L=\epsilon L_0$, $\delta \alpha=\epsilon \beta \omega $, and $\dt E= \epsilon^2 E_0$.

We find  that 
\begin{align}  
 \psi_{\ell m} (\theta,\phi)= c_{\ell m} {\rm Re}({\rm Y}_{\ell m})+ b \cos \st
\end{align} 
solves Eq. \eqref{zeromode}, if 
\bea 
\beta=\beta_{c,\ell}=-\frac{\ell(\ell+1)}{ 2n_0},
\eea 
and $b=-n_0 \beta \omega/(1+n_0 \beta)$
for $\ell \neq 1$.  While for $\ell=1$,  $b$ can be  arbitrary  and we choose that $b=0$ without losing generality.  
Here ${\rm Y}_{\ell m}$ is the spherical harmonic function. 
For weakly clustered states, the vorticity field $\sigma=\sigma_{\ell m}\equiv \delta \sigma=\epsilon \ell(\ell+1)\psi_{\ell m}$.  

%\[ a=\begin{cases} 
%	-\frac{n_0 \beta \omega}{1+n_0 \beta} & \ell \neq1\\
%	\omega=0 & \ell=1
%\end{cases}
%\]

Given the values of $E_0$ and $L_0$,  $c_{\ell m}$ and $\omega$ are determined by 
\begin{align}
   E_0&= - \frac 12 \int  \d \Omega \,  \psi_{\ell m}(\theta, \phi)\nabla^2 \psi_{\ell m}(\theta, \phi),\label{E0} \\
   L_0&=- 2n_0 \beta_{c,\ell} \int \d \Omega  \cos \st \ \psi_{\ell m} -\frac{2}{3} \beta_{c,\ell} \omega. 
    \label{L0}  
\end{align} 
For $\ell=m=0$, $E_0=0$ and $L_0=0$, hence only modes with $\ell>0$ are relevant.  In the following we focus on rotation free states, namely $\omega=\alpha=\delta \alpha =0$, and in this case $\psi_{\ell m}=c_{\ell m} \rm{Re} ({\rm Y}_{\ell m}) $.

\subsubsection{dipole states: vortex clustered states with finite angular momentum} \label{diplestate}
 Let us firstly consider the state with $\ell=1$ and $m=0$, the corresponding stream function is   
\bea
\psi_{10}=c_{10}{\rm Y}_{10}=c_{10}\sqrt{\frac3{4\pi}}\cos \st,
\eea
where $ c_{10}=\sqrt{E_0}$.  This state carries finite angular momentum and $L_0\neq 0$. 

The dipole moment defined as 
\begin{align} 
	\bm D \equiv \int \d \Omega \ \sigma(\bm r) \, \bm r,  
\end{align}  
and for this state the dipole moment reads
\bea
D_{\eta}= D_{\xi}=0 \quad \text{and} \quad D_{\zeta}=4\sqrt{\frac{\pi}{3}}\sqrt{E}.
\label{dipolemoment}
\eea
We refer to this state as {\em dipole state} and
the onset of the dipole state  occurs at $\beta=\beta_{c,1}\equiv \beta^d_c=-4\pi$. For the dipole state, the angular momentum $L_{\zeta}=4\sqrt{\pi/3}\sqrt{E}$.
It is interesting to  note that the vorticity distribution for the dipole state at low energies is identical to the vorticity distribution for a stationary vortex flow on a sphere~\cite{Xiong2024}, obtained from the vortex fluid theory.

For $ m=\pm 1$, the corresponding stream functions  are  
$\psi_{11}=-\psi_{1,-1}=c_{11} {\rm Re} ({\rm Y}_{11})= -c_{11}\sqrt{3/8\pi} \sin \theta \cos \phi$,
where $ c_{11} =\sqrt{2}c_{10} $.  It happens that these modes are connected by a $\textrm{SO}(3)$ rotation, namely, $\psi_{11}(\bm{r})=\psi_{10}(\bm{r}_1)$ and $\bm{r}_1=R (0,\pi/2, 0 ) \, \bm{r}$.  Here $R(\theta_1, \theta_2, \theta_3)=R_{\zeta} (\theta_3) R_{\eta}(\theta_2) R_{\zeta}(\theta_1)$ is the $\textrm{SO}(3)$ rotation operator and $\{\theta_{i=1,2,3}\}$ are the Euler angles.  Therefore modes $\psi_{10}$, $\psi_{11}$ and $\psi_{1-1}$ describe the same dipole state while with different dipole moment orientations. Hereafter we focus on the dipole state described by the mode $\psi_{10}$ for which the dipole moment is along the $\zeta$-axis  [Fig.~\ref{QS} (a)] .

\subsubsection{Sandwich states: clusters with zero angular momentum}
\label{sandwichstate}
In this subsection, we consider clustered states with zero angular momentum $\bm{L} =0$ (or $L_0=0$). Hence only modes with $\ell>1$  are relevant and  $\delta L_{\zeta}=\delta \alpha =0$.  Intuitively modes with larger $\ell$ describe richer and finer structure and should have lower entropy. To ensure it is the case,  let us  
expand $S[n_0 +\delta n_-,n_0+\delta n_+]$ around the uniform density $n_0$ up to second order in $\delta n_{\pm}$, and we  obtain
\bea
S&=& S_0-\frac{1}{2n_0} \int \d\Omega \left[(\delta n_+)^2+(\delta n_-)^2\right]+\smallO(\delta n_{\pm}) \nn\\
&=&S_0-\frac{1}{2n_0} \left[\left(\int \d\Omega\ {\rm e}^{-\beta \delta \psi}\right)^{-2}  \int \d\Omega\, {\rm e}^{-2\beta \delta \psi} \nn \right. \\ 
&&\left .+\left(\int \d\Omega\, {\rm e}^{\beta \delta \psi}\right)^{-2}  \int \d\Omega \, { \rm  e} ^{2\beta \delta \psi}-2n_0\right]+\smallO(\delta n_{\pm}),
\label{entropychange}
\eea  
where  the first order term in $\delta n_{\pm}$ vanishes due to the constraint Eq.~\eqref{c1} and $S_0 =- \int \d\Omega \ 2n_0 \log n_0$ is the entropy for the homogeneous state. Here we have used $\delta n_{\pm}=n_0 [{\rm e}^{\delta \gamma_{\pm}} {\rm e}^{\mp \beta \delta \psi}-1]$ and  Eq.~\eqref{c11} for $\delta \alpha=0$. To leading order in $\epsilon$, $\delta n_{\pm}\simeq \mp \epsilon n_0 \beta_{c,\ell} \psi_{\ell, m} $ and $\delta \gamma_{\pm}=0$, then  Eq.~\eqref{entropychange} becomes 
\bea
S&&\simeq S_0-\left[n_0\beta^2_{c,\ell} \int \d\Omega \, \psi^2_{\ell m} \right]\epsilon^2 \nn\\
&&=S_0-\frac{1}{E_0}\left[n_0\beta^2_{c,\ell} \int \d \Omega \, \psi^2_{\ell m} \right] \delta E,
\label{entropychangeleadingorder}
\eea   
and one can easily show that modes with $\ell =2$ have higher statistical weight (larger entropy) than the modes with $\ell >2$ (see Sec. \ref{A} in Appendix).

\begin{figure}[htbp]
\centering  
\includegraphics[width=8.9cm]{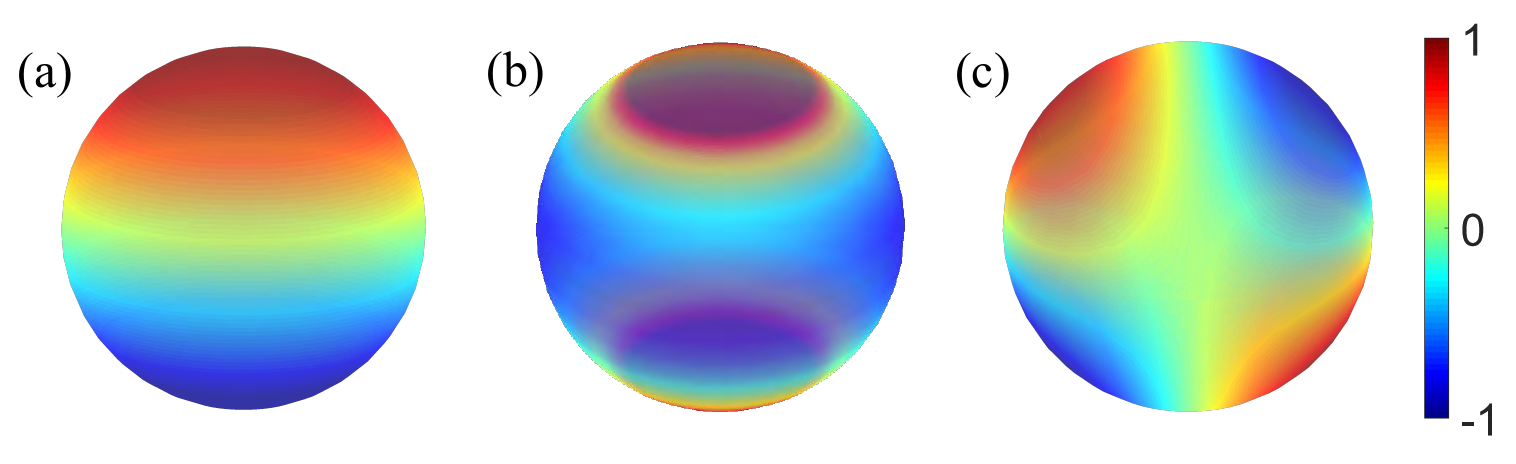}
\caption{ Vorticity $\sigma$ for the dipole (a), the sandwich (b) and the quadrupole (c) state. The extreme values of $\sigma$ are scaled to unity for convenience.  \label{QS}}
\end{figure}

Hence we only need to focus on states with $\ell=2$. For $m=0$,  
\bea
\psi_{20}=c_{20} {\rm Y}_{20}=\frac{c_{20}}{4} \sqrt{\frac5\pi}(3\cos^2\st-1), 
\label{streamfunctionsandwich}
\eea
where $c_{20}=\sqrt{E_0/3}$.  This mode describes the clustering of positive (negative) vortices around the poles and negative (positive) vortices around the  equator [Fig.\ref{QS} (b)].  We refer to this state as {\em sandwich state}.  The onset of the sandwich state occurs at $\beta=\beta_{c,2}\equiv \beta_c=-12 \pi$.  For $m=1$ and $m=2$, $\psi_{22}(\theta,\phi)=c_{22}{\rm Re}({\rm Y}_{22})=(c_{22}/4)\sqrt{15/2\pi}\cos (2\phi) \sin^2\st$, $\psi_{21}(\theta,\phi)=c_{21}{\rm Re}({\rm Y}_{21})=-(c_{21}/4)\sqrt{15/2\pi}\cos \phi \sin2\st$, where $c_{22}=c_{21}=\sqrt{2E_0/3}$. Here the two modes are also connected by a $\textrm{SO}(3)$ rotation, i.e., $\psi_{21}(\bm{r}_2)=\psi_{22}(\bm{r})$, where $\bm{r}_2=R (-\pi/4,\pi/2,\pi/2) \, \bm{r}$.  Therefore, modes $\psi_{22}$ and $\psi_{21}$  describe the same state and we refer to this state as {\em quadrupole state}.  Note that $\psi_{2,-2}=\psi_{2,2}$ and $\psi_{2,-1}=-\psi_{2,1}$, hereafter we focus on the quadrupole state described by the mode $\psi_{21}$[Fig.\ref{QS}(c)].

For states with $\ell=2$,  the dipole moment $\bm D$ vanishes.  In order to quantify the clustered states with zero angular momentum, it is necessary to introduce the quadrupole tensor 
\begin{align}
 	\mathcal{Q}_{ij}\left[\sigma(\bm r) \right] \equiv \frac12 \int \d\Omega \, \Big[3(\bm r\cdot \bm e_i)(\bm r\cdot \bm e_j)-\dt_{ij} \Big]\s(\bm r),
 \end{align}  
where $\{\bm e_{i=\xi,\eta,\zeta}\}$ are unit vectors in Cartesian coordinates. 
The quadrupole moment is defined as 
 \bea
 Q[\sigma] \equiv \sqrt{q^2_1+q^2_2+q^2_3}, \label{QM}
 \eea
where $\{q_1,q_2,q_3\}$ are eigenvalues of $\mathcal{Q}_{ij}$. We denote $Q_s$ and $Q_q$ as the quadrupole moment for the sandwich and the quadrupole state, respectively.  

For the sandwich state, 
 \bea
  \mathcal Q[\sigma_{20}]= \epsilon c_{20} \sqrt{\frac{36 \pi}{5} }\left( 
 \begin{array}{ccc}   
 	-1 & 0 & 0\\  
 	0& -1 & 0\\  
 	0& 0 &2\\
 \end{array}
 \right)   
 \eea
and its eigenvalues are $\epsilon c_{20} \sqrt{36 \pi/5 }\, \{-1,-1,2\}$. The corresponding quadrupole momentum is 
\bea 
 Q_s=Q[\sigma_{20}]=Q_0 |E-E_c|^{\nu},
\label{quadrupolemoment}
\eea
where $\nu=1/2$, $Q_0=\sqrt{216\pi/15}$ and $E_c=0$.
For the quadrupole state, 
 \bea       
 \mathcal Q[\sigma_{21}]&&=-\epsilon {c_{21}} \sqrt{\frac{54\pi}{ 5}}\left(   
 \begin{array}{ccc}  
 	0& 0 & 1\\  
 	0& 0 & 0\\  
 	1& 0 &0\\
 \end{array}
 \right)  {,}\ 
 \eea
% and
% \bea   
% \mathcal Q [\sigma_{22}]&&= \epsilon {c_{22}}\sqrt{\frac{54\pi}{ 5}}\left(                 
% \begin{array}{ccc}    
% 	1& 0 & 0\\  
% 	0 & -1 & 0\\   
% 	0&0&0\\
% \end{array}
% \right).
% \eea
and the eigenvalues of $\mathcal Q[\sigma_{21}]$ are $ - {c_{21}} \epsilon\sqrt{54\pi/ 5}\, \{-1,0,1\}$. The corresponding quadrupole momentum  takes the same value as for the sandwich state, namely,  $Q_q=Q_s$.
Note that since $\sigma_{21}(R \bm{r})=\sigma_{22}(\bm{r})$, it is easy to see that  $\mathcal{Q}[\sigma_{21}]=R\mathcal{Q}[\sigma_{22}]R^{T}$, where $R=R (-\pi/4,\pi/2,\pi/2)$.

Applying Eq.~\eqref{entropychangeleadingorder}, we find that up to order $\delta E$, the sandwich state and the quadrupole state have the same entropy and the entropy difference appears in order $\delta E^2$:
\bea 
S[ {\rm sandwich } ] -S[ {\rm quadrupole} ] \simeq \frac{5 \beta_c^4 E_0^2 \epsilon ^4}{576 \pi ^2}=\frac{5 \beta_c ^4}{576 \pi ^2} (\delta E)^2. \nn\\
\eea
Hence under condition of zero angular momentum, \emph{ the sandwich state has the highest statistical weight}.  Although this conclusion is made based on the perturbative analysis at low energies, it will be confirmed by later analysis at high energies and Monte Carlo simulations at moderate energies.

%\begin{figure}[htp]
%	\centering
%	\includegraphics[width=3.5cm]{Y10.png} 
%	\caption{Vorticity density of dipole configuration.}
%	\label{Y10}
%\end{figure} 

 \subsection{High energy configurations}
In this subsection, we consider the sandwich state and the dipole state  at high energies.  In the high energy limit, vortices and antivortices are well separated and are concentrated in small regions of a sphere. 
It is then convenient to  use the stereographic projection method to find approximate density distributions of  vortices on a sphere.

In stereographic coordinates  Eq.~\eqref{selfconsistent} becomes
 \begin{align}  
 \ (1+z \bar z)^2  \p_{z }\p_{\bar z} \psi ={\rm e}^{\beta \psi(z ,\bar z)+\gamma_-}-{\rm e}^{-\beta \psi(z ,\bar z)+\gamma_+}  
 \label{Poissonz} 
 \end{align}
for rotation free states ($\omega=\alpha=0$).
The stereographic projection maps the southern hemisphere onto a disc on the $\zeta=0$ plane (Fig.\ref{proj}) and transforms vortices in the region around southern pole and vortices in the region around the equator to the region around the origin and the region near the boundary of the disc, respectively.

\subsubsection{Sandwich states at high energies} 
In the high energy limit, the interaction between vortices near the poles and antivortices near the equator can be neglected.  For $|z|\ll1$, Eq.~\eqref{Poissonz} is well approximated as 
\begin{align} 
  \p_{z }\p_{\bar z} \psi = -{\rm e}^{-\beta \psi(z ,\bar z)+\gamma_+},
 \label{Poissonz1} 
 \end{align}  
 which can be regarded as the mean-field equation of describing distributions of chiral vortices confined to a (flat) disc and  is well-defined on the whole disc domain. Hence finding the approximated distribution of vortices near the poles at high energies is boiled down to searching for distributions of chiral vortices confined to a disc with proper boundary conditions. 
 
For axisymmetric distributions, Eq.~\eqref{Poissonz1} becomes 
\begin{align} 
    \frac{1}{4} \frac{1}{\rho} \frac{\d}{\d\rho} \left( \rho \frac{\d}{\d \rho}\psi \right)=- {\rm e}^{- \beta \psi(\rho)+\gamma_+},
    \label{poissd1}
\end{align} 
and has the following exact solution	
\begin{align}
\psi(\rho)&=-\frac{2}{\beta} \log  \left(\frac{2}{\beta \rho^2 {\rm e}^{\gamma_+} -2  }\right)
\label{psic}
\end{align}  
with boundary condition $\psi(\rho=0)=\psi'(\rho)|_{\rho=0}=0$~\cite{ SmithONeil, Kiessling1995,Patil2021,Xiong2023iop}.  Here the boundary condition $\psi'(\rho)|_{\rho=0}=0$ for the problem of chiral vortices confined to a disc is induced by the flow generated by the vortices on the sphere and can be obtained from the global stream function Eq.~\eqref{streamfunctionsandwich} at low energies. Also the  vortex density distribution Eq.~\eqref{nc} does not depend on the value of $\psi(\rho=0)$  and here we choose $\psi(\rho=0)=0$ for convenience [see Sec. \ref{b} in Appendix].
We emphasize again that the exact solution Eq.~\eqref{psic}  is valid in the whole disc while provides a good approximation of the distribution of vortices near the poles at large energies.  

The vortex density then reads 
\bea
n_+(\rho)=\exp(-\beta \psi+\gamma_+ )= \frac{4 {\rm e}^{\gamma_+} }{ (  \beta \rho^2  {\rm e}^{\gamma_+} -2)^2}.
\label{nc}
\eea   
For the sandwich state, since only half positive vortices are distributed around the south pole (the other half positive vortices are distributed around the north pole), the normalization condition is
\begin{align}
 	\frac{1}{2} = \int^{2\pi}_0  \d \phi  \int_0^1  \frac{4\rho}{(1+\rho^2)^2} \, \d \rho \, n_+(\rho), \label{N1} 
\end{align}  
which gives rise to the relation between ${\rm e}^{\gamma_+}$ and $\beta$ [Fig.~\ref{betag}(a) in Appendix].

When $\beta \rightarrow \beta_s= -16\pi$,  vortices around the pole collapse into a point, known as the supercondensation. 
Hence the parameter range for the sandwich state is $\beta_s<\beta<\beta_c$.  The limit configuration at $\beta=\beta_s$ is $ n_+(\theta)= [\dt(\theta)+\dt(\theta-\pi)]/(4 \pi \sin\theta)$.  It is worthwhile mentioning that the value of $\beta_s$ depends on the normalization condition. For vortices confined to a disc on a plane, the vortex density profile is the same as Eq.~\eqref{nc}, however the vortex number is normalized to unity, which gives rise to that $\beta_s=-8\pi$ (in convention of this paper)~\cite{ SmithONeil, clusteringYu, Kiessling1995,  Patil2021, Xiong2023iop}. 

\begin{figure}[htp]  
    \centering
    \includegraphics[width=4.1cm]{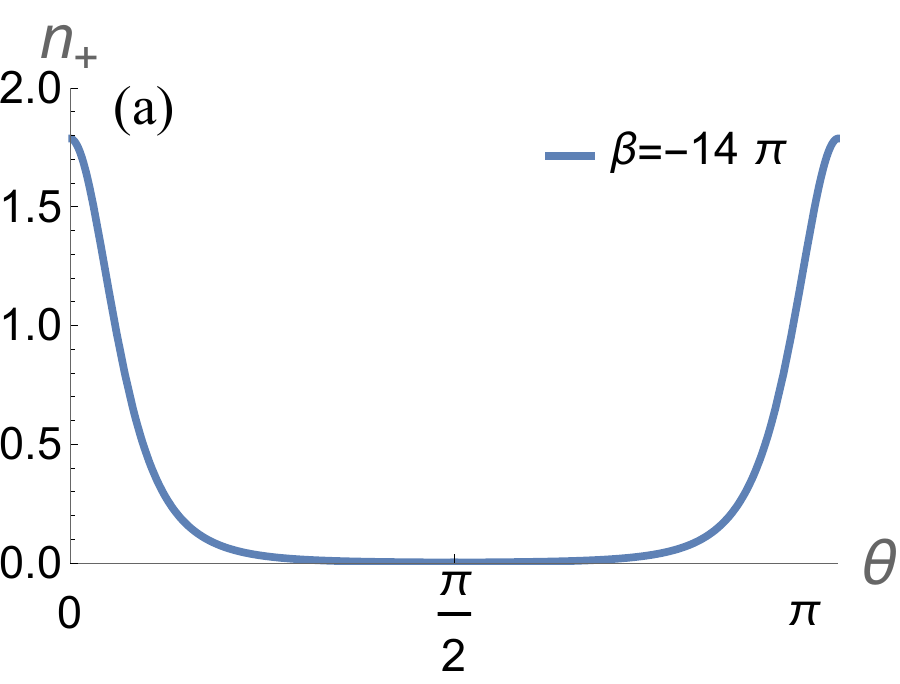}
   \includegraphics[width=4.1cm]{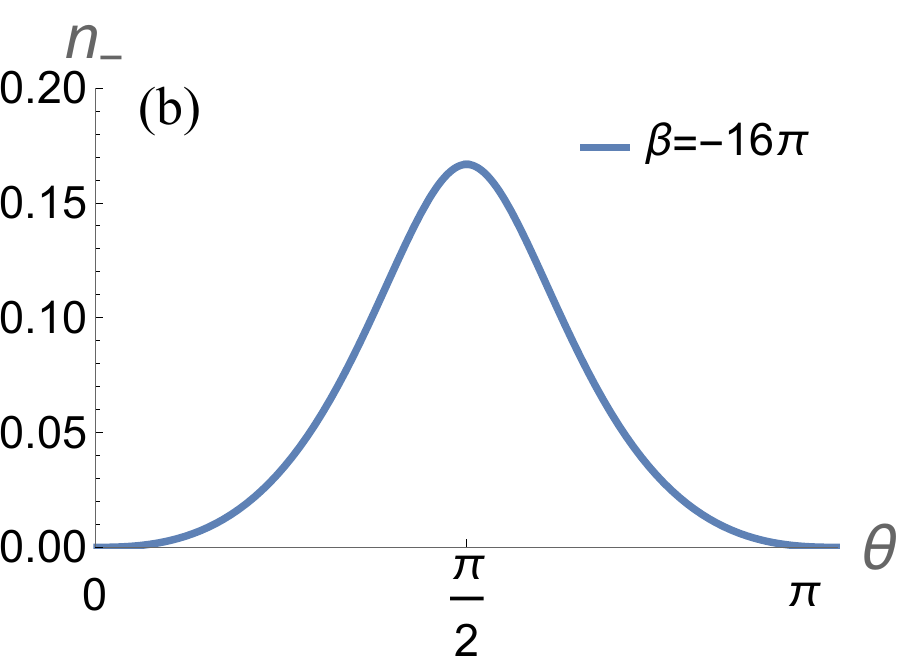} 
    \caption{The vortex density distributions for the sandwich state at high energies. (a) The distribution of positive vortices near poles [Eq.~\eqref{sandp}] . (b) The distribution of negative vortices near the equator at $\beta=\beta_s$ [Eq.~\eqref{sande}].}
    \label{Sandpole}
\end{figure}

In spherical coordinates,  the stream function and the vortices density on the southern hemispheres [$\st \in (\pi/2,\pi)$] are [see also Fig.~\ref{Sandpole}(a)]
\bea
	\label{streamfunctiondipole}
	\psi(\theta)&&=-\frac{2}{\beta} \log  \left[\frac{2}{ \beta  \cot^2 \left(\frac{\theta}{2}\right) {\rm e}^{\gamma_+} -2  }\right],\\
   n_+(\theta)&&=\frac{4 {\rm e}^{\gamma_+} }{ \left [\beta \cot^2\left(\frac{\st}{2}\right) {\rm e}^{\gamma_+}   -2 \right ]^2}.\label{sandp}  
\eea

Now we consider the distribution of negative vortices in the region near the equator which is mapped to the region near the edge of the disc.  For $|z| \sim 1$, Eq.~\eqref{Poissonz} is well approximated as 
\begin{align} 
 4\p_{z }\p_{\bar z} \psi &={\rm e}^{\beta \psi(z,\bar z) +\gamma_-}.
  \label{Poissonz3}
\end{align}  
For axis-symmetric solutions, Eq.~\eqref{Poissonz3} becomes
\bea
 \frac{1}{\rho} \frac{\d}{\d \rho}\left ( \rho \frac{\d}{\d \rho}\psi \right)&={\rm e}^{\beta \psi(\rho)+\gamma_-}, \label{Poissiond2}
 \eea  
which admits an exact solution
 \begin{align} 
 \psi(\rho)=\frac 2 \beta \log \left(\frac{2A \rho^{(-1 + A/2 ) }}   {2 + A+ (A -2 ) \rho^{A } }\right) ,\label{psiedge}
  \end{align} 
with the boundary condition~\cite{ Xiong2023iop} 
 \bea
 \psi(\rho=1)=\psi'(\rho)|_{\rho=1}=0
 \label{boundarycondition}
 \eea
and $A=\pm \sqrt{4-2 \beta {\rm e}^{\gamma_-}  }$. 
Similarly, the exact solution Eq.~\eqref{psiedge} to the problem of chiral  vortex clusters in a disc with the boundary condition Eq.~\eqref{boundarycondition}  is  well-defined on the whole disc domain and provides a good approximation of the distribution of vortices near the equator at large energies.

As it follows, the density is 
\bea
  n_-(\rho)=\frac{4  A^2 \rho^{( A-2 )}{\rm e}^{\gamma_-} }{ \left [ 2 + A+ (A -2 ) \rho^{A }  \right ]^2}.
  \label{ned}
\eea  
Note that  the boundary condition $\psi'(\rho)|_{\rho=1}=0$ is also induced by the global stream function Eq.~\eqref{streamfunctionsandwich}.  The  vortex density distribution Eq.~\eqref{ned}  is independent of the value of $\psi(\rho=1)$ and here we choose $\psi(\rho=1)=0$ for convenience [see Sec. \ref{b} in Appendix].
The relation between ${\rm e}^{\gamma_-}$ and $\beta$ is given by the normalization condition Eq.~\eqref{N1} ($n_{+} \rightarrow n_{-}$).  Moreover, the smooth condition $n'_{-}(\rho)|_{\rho=0}=0$ requires ${\rm e}^{\gamma_-}\beta<-5/2$  which gives rise to that $\beta<-7.11 \pi$ [Fig.~\ref{betag} (b) in Appendix]. Hence for $\beta$ within the range $\beta_s<\beta<\beta_c$, the smooth condition is automatically satisfied.

In spherical coordinates, the stream function and the vortex density distribution of negative vortices near the equator [$\st \in (\pi/2,\pi)$] read 
 \begin{align} 
 \psi(\theta)&=\frac 2 \beta \log \left[\frac{2A \cot (\theta/2)^{(-1 + A/2 ) }}   {2 + A+ (A -2 ) \cot (\theta/2)^{A } }\right], \\    
  n_-(\theta)&=\frac{4 A^2 \cot^{( A -2)}(\st/2) {\rm e}^{\gamma_-} }{ \left [2 + A+ ( A -2 ) \cot^{A }(\st/2)  \right ]^2} . \label{sande}
 \end{align}  
Induced by the boundary condition Eq.~\eqref{boundarycondition}, the fluid velocity $ u_\phi=-\partial \psi (\theta) /\partial \theta$ on the equator vanishes, namely  $u_\phi(\theta=\pi/2)=0$, which is consistent with the fluid velocity field on the equator at low energies [$\psi_{20}'(\theta)|_{\theta=\pi/2}=0$]. 
At $\beta=\beta_s$,  ${\rm e}^{\gamma_-}\approx 0.167,   A\approx 4.556$ [Fig.~\ref{betag}(b) in Appendix], the vorticity for the sandwich state reads
$\sigma_s(\bm r)=\Big[\dt(\theta)+\dt(\theta-\pi) \Big] \Big/(4\pi \sin\theta)-n_-(\theta), 
$
and the quadrupole moment reaches the maximum value $Q^{\rm max}_s \approx 1.59 $. 
It is easy to see that the  quadrupole state also reaches the high energy limit at $\beta=\beta_s$,  
and the corresponding vorticity is 
$\s_{q}(\bm r)=\Big[ \dt(\theta-\pi/4,\phi)+\dt(\theta-3\pi/4,\phi-\pi)   - \dt(\theta-3\pi/4,\phi) -\dt(\theta-\pi/4,\phi-\pi) \Big]/(2\sin\theta) $
which gives rise to the maximum quadrupole moment  $Q^{\rm max}_{q}=3\sqrt{2}/2>Q^{\rm max}_s$.  Note that  for the quadrupole state  locations of the clusters at supercondensation coincide with mechanical equilibrium positions for $4$ point vortices on a sphere.

\subsubsection{General solutions}
A general solution to Eq.~\eqref{Poissonz1} is available~\cite{CROWDY1997,henrici1993} :
\begin{align}
    \psi(z,\bar z) =& -\frac{1}{\beta}\log\left[\frac{2f'(z) \overline{f'(z)}}{-c \beta[1+f(z) \overline{f(z)}]^2} \right ], \label{gs}\\
n(z,\bar z)=& c \exp(-\beta \psi),
\label{general}
\end{align}
where $f(z)$ is a meromorphic function of $z$ in a simply connected domain and $\overline {f(z)}$ is the complex conjugate of $f(z)$. Furthermore, it is required that  $f(z)$ has at most simple poles and $f'(z) \neq 0$ in the consider domain~\cite{henrici1993,stuart2004}.

Here we show that Eqs.~\eqref{psic} and~\eqref{psiedge} fall into this general solution by specifying function $f(z)$ properly.   For an axisymmetric distribution, the stream function $\psi(z,\bar{z})$ must be the function of $z \bar z$ or $\rho$ only.
Hence $f(z)$ has a unique form and $f(z)=a z^{b}$, here $a \in \mathbb C, b \in \mathbb Z$. 
The boundary condition $\psi(\rho=0)=\psi'(\rho)|_{\rho=0}=0$ determines  that $a=\sqrt{-c \beta /2}$, $b =1$, and $c={\rm e}^{\gamma_+}$, which leads to Eq.~\eqref{psic} precisely.  The general solution Eq.~\eqref{gs} gives rise to Eq.~\eqref{psiedge} by choosing $ f(z)=\  \sqrt{ (A-2)/(A+2)}\ z^{A/2}$, where $ A=\pm \sqrt{4-2 \beta {\rm e}^{\gamma_-} }$, and $ c ={\rm e}^{\gamma_-}/4$ . Here the boundary condition $\psi(\rho=1)=\psi'(\rho)|_{\rho=1}=0$ is fulfilled. An observation can be made here:  the exponent $A/2$ is \emph {not} an integer in general. Hence such a choice of $f(z)$ may not be a meromorphic function and then does not meet the  requirements for  Eq.~\eqref{gs}.  As shown previously,  such a solution is indeed the solution to Eq.~\eqref{Poissiond2}, suggesting that the validity condition of Eq.~\eqref{general} may  be extended.  

 \subsubsection{Dipole states at high energies} 
At high energies,  for the dipole state, vortices with opposite sign are concentrated near the poles.  The density distribution of positive vortices near the southern pole is identical to Eq.~\eqref{nc}
and the only difference is that here the vortex density is normalized to unity but not 1/2.  Hence for the dipole state, the supercondensation occurs at $\beta=\beta^d_s=-8 \pi$ which is the same as what for vortices confined to a disc on a plane~\cite{SmithONeil, clusteringYu, Xiong2023iop}.

\begin{figure}[htp]
	\centering
	\includegraphics[width=6cm]{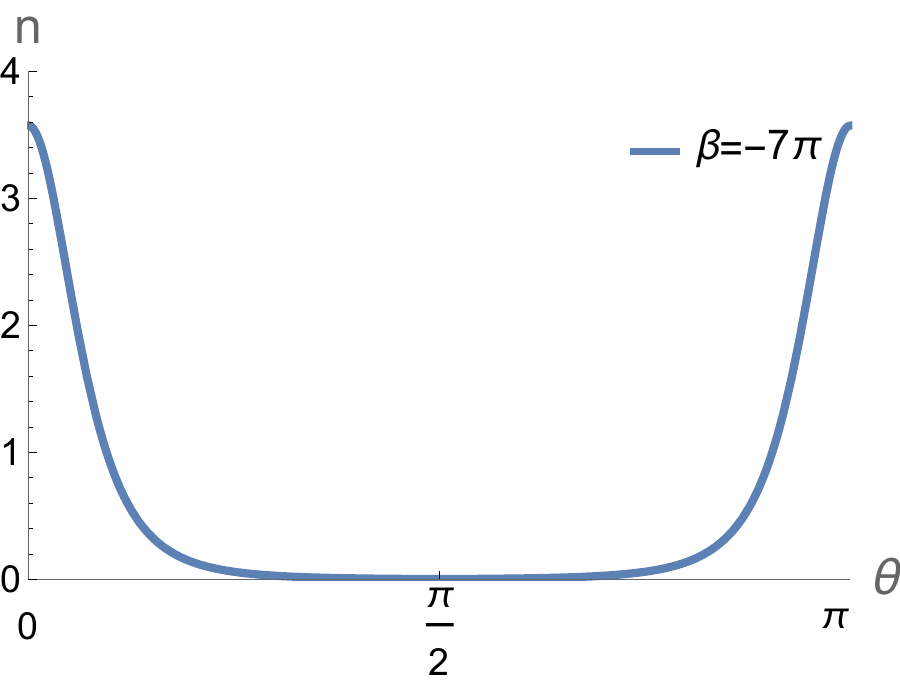}
	\caption{The total vortex density for the dipole state at high energies  [Eqs.~\eqref{DipSph1} and \eqref{DipSph2}]. }
	\label{dipole}
\end{figure}

In spherical coordinates, vortex densities are [see also Fig.~\ref{dipole}]  
\bea
n_-(\st)&&= \frac{4 {\rm e}^{\gamma_-}}{\Big[\beta \cot ^2(\st/2) {\rm e}^{\gamma_-}-2 \Big]^2  } \quad \st \in (\pi/2,\pi),\label{DipSph1}\\ 
n_+(\st)&&= \frac{4{\rm e}^{\gamma_+} }{\Big[ \beta \tan^2( \st /2) {\rm e}^{\gamma_+  } -2\Big]^2 } \quad \st \in (0,\pi/2),
\label{DipSph2}
\eea  
and the corresponding angular momentum is 
\begin{align} 
	L_{\zeta}(\beta)=& \frac{ 8 \pi {\rm e}^{\gamma_-}}{ (2+  \beta {\rm e}^{\gamma_-})^4} \Big[  \beta {\rm e}^{\gamma_-}\Big( \beta {\rm e}^{\gamma_-}(-2 \beta+5+32 \log 2) \\
	&+20- 32\log 2 \Big) 
	-16 \beta {\rm e}^{\gamma_-}( \beta {\rm e}^{\gamma_-}-1) \log(2- \beta {\rm e}^{\gamma_-})+ 4 \Big]. \nn
\end{align}  
When $\beta \rightarrow \beta^d_s$, $n_-\rightarrow \delta(\theta-\pi)/(2\pi \sin\theta),n_+\rightarrow \delta(\theta)/(2\pi \sin\theta)$, and the angular momentum reaches the upper bound $L_{\zeta} \rightarrow L^{\rm{max}}_\zeta=2$. 

\begin{figure}[htbp]
	\centering
	\includegraphics[width=6cm]{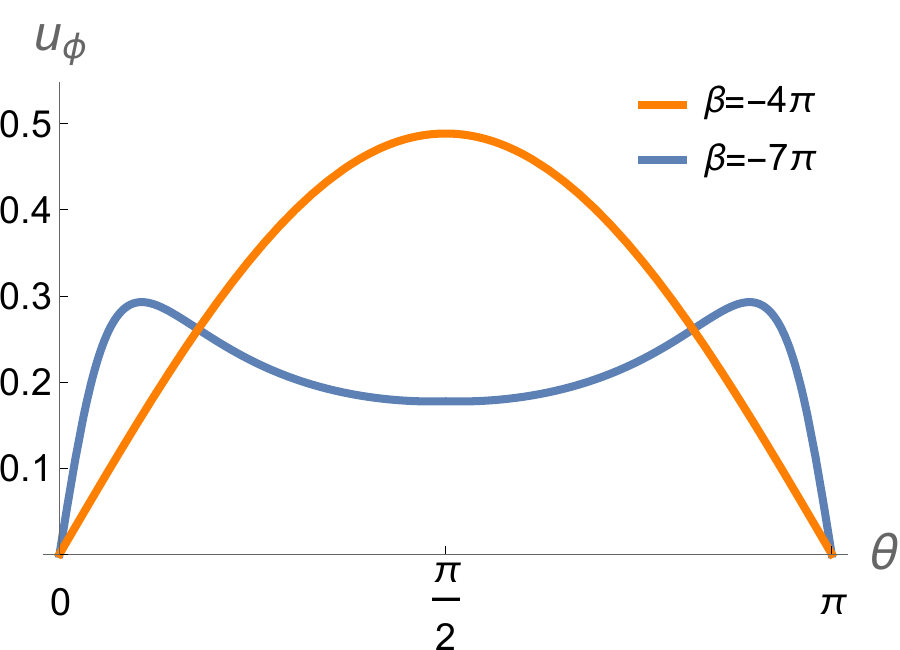}
	\caption{The velocity field $ u_\phi$ for the dipole state  at a low energy (orange solid line) and at a high energy (blue solid line). The parameter region for the dipole state is $-8\pi=\beta^d_s<\beta<\beta^d_c=-4\pi$.   When $\beta \to \beta^d_s$, $u_\phi(\theta=\pi/2)\rightarrow 1/2\pi$.  Note that here the fluid velocity field on the northern hemisphere is obtained by  shifting the coordinate of Eq.~\eqref{vh} ($\theta \rightarrow \pi-\theta $). The two expressions  work well only near the poles and  take different values on the equator.  However  the difference decreases as increasing energy and for the $\beta$ value we choose here, the discrepancy is invisible.   
	}
	\label{velocitydip}
\end{figure}

We now consider the flow generated by these vortices. The fluid velocity field reads   $\bm{u}=\nabla \psi \times \bm{e}_r=u_{\theta} \bm{e}_{\theta}+ u_{\phi} \bm{e}_{\phi}$,  where $\bm{e}_r$, $\bm{e}_{\theta}$, $\bm{e}_{\phi}$ are three  orthogonal spherical unit vectors.  For the dipole state at low energies, $u_\theta=(1/\sin \theta) \partial \psi/\partial \phi=0$, $u_\phi=- \partial \psi /\partial \theta \propto \sin \theta $ and  $u_\phi$ vanishes at poles (Fig. \ref{velocitydip}).  Here 
At high energies,  the velocity field is 
\begin{align}
	u_\phi(\theta)= - \frac{\partial \psi }{\partial \theta}=\frac{2 \cot(\theta/2) \gamma_- }{[ 2 - \beta \gamma_- \cot^2(\theta/2)]\sin^2(\theta/2)},
	\label{vh}
\end{align}
where we have used Eq.~\eqref{streamfunctiondipole}.
Consistently, it also  vanishes at poles.  
At low energies, fluid velocity field $u_\phi$ reaches the peak value on the equator ($\theta=\pi/2$).  While at high energies, two peaks of the velocity field  start to develop and move towards to poles  as increasing the energy  (Fig.~\ref{velocitydip}). This is because for tightly clustered vortices around the poles,  the fluid velocity is considerably large on the edges of the vortex clusters which locate between the poles and the equator.  

\subsection{Comparison of the entropy  between the sandwich state and the quadrupole state at high energies}

\begin{figure}[htp]
	\centering
	\includegraphics[width=7.5cm]{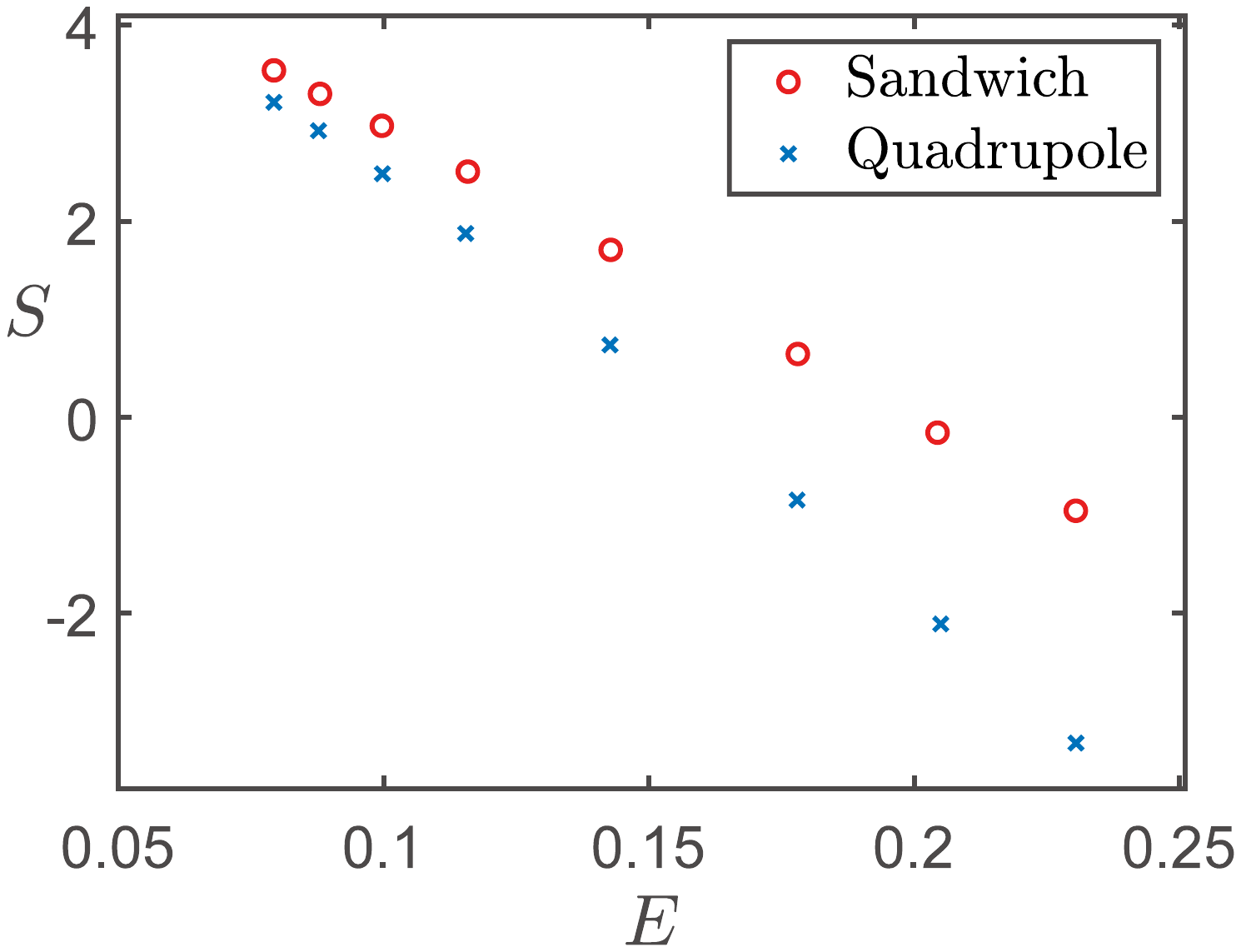}
	\caption{A comparison between the entropy of the sandwich state (red circle) and the quadrupole state (blue cross) at high energies. The entropy values are evaluated using Eqs.~\eqref{sandp} and ~\eqref{sande} for the sandwich state and  Eqs.~\eqref{DipSph1} and \eqref{DipSph2} for the quadrupole state. }  
	\label{entropyhighenergy}
\end{figure}

The quadrupole state at high energies can be easily constructed from Eqs.~\eqref{DipSph1} and ~\eqref{DipSph2}.  From the approximate high energy expressions of vortex densities,  we can calculate the entropy and find again that  $S[\text{sandwich state}]> S[\text{quadrupole state}]$ (Fig.~\ref{entropyhighenergy}).

\subsection {Gaussian clustered states}
In this section, we discuss a  limit distribution  when $\beta \rightarrow0$, $\omega \rightarrow \infty $ simultaneously while $\alpha=\beta \omega$ is finite.  In this limit, Eq.~\eqref{n2} admits an exact solution: 
 \begin{align}
	n_\pm(\theta) = &\frac{\alpha}{4\pi \sinh\alpha} \exp(\mp \alpha \cos\theta ), \\
	\psi(\theta) = & -\frac{1}{8\pi \sinh \alpha} \Big [ {\rm e}^{\alpha} {\rm Ei}\Big(\alpha(\cos\theta-1)\Big)-{\rm e}^{\alpha}{\rm Ei}\Big(-\alpha(\cos\theta+1)\Big )  \notag\\   
	& + {\rm e}^{-\alpha}{\rm Ei}\Big(\alpha(1-\cos\theta) \Big)-{\rm e}^{-\alpha} {\rm Ei} \Big(\alpha(\cos\theta+1)\Big) \nn\\
	&  -2 \alpha C \ \rm{arctanh}(\cos \theta)\Big] ,
	\label{Gaussstream}
\end{align}  
where $\alpha \in (-\infty,\infty)$,  ${\rm Ei}(x)=\int^x_{-\infty}{\rm e}^{t}/t\  \d t$, and $C$ is a constant determined by proper smooth conditions. 
The angular momentum for this Gaussian state is
\begin{align}
	 L_{\zeta}= \int (n_+-n_-)  \cos\theta\, \d \Omega=\frac{2}{\alpha }  (1-\alpha \coth{\alpha} ) .  
\end{align}     

\begin{figure}[htp]
  \centering
    \includegraphics[width=4.3cm]{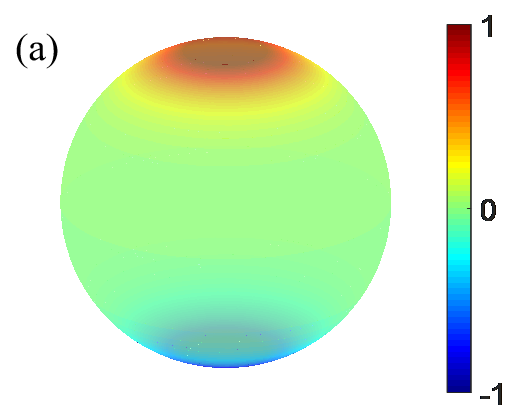}\\
    \includegraphics[width=4.1cm]{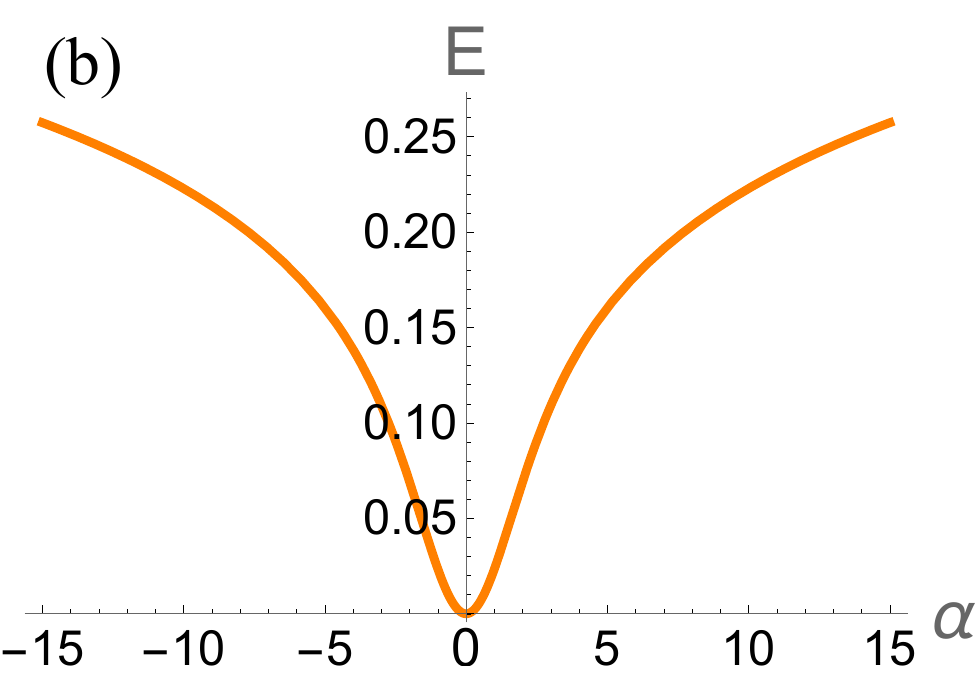}
    \includegraphics[width=4.1cm]{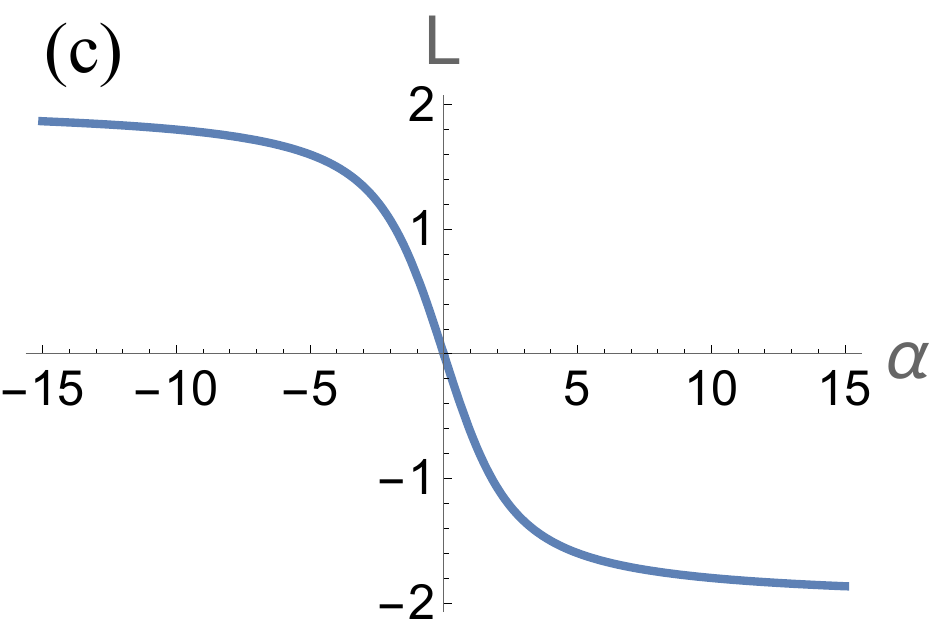}
    \caption{(a) Vorticity $\sigma=n_+-n_-$ for the Gaussian state at $\alpha=-8$. (b) and (c) show the energy and the angular momentum of as functions of $\alpha$, respectively.  }
    \label{Gauss}
 \end{figure}

The components of the fluid velocity field for this state are  $u_\theta=0$ and

\begin{align}
	u_\phi=-\frac{\partial \psi}{\partial \theta} = \frac{ \alpha}{4\pi \sinh \alpha \sin \theta}  \left[ \frac{2\cosh({\alpha \cos \theta})}{\alpha}+C \right] .
\end{align}

Here we impose the zero velocity condition at poles, namely  $u_\phi(\theta=0)=u_\phi(\theta=\pi)=0$,  giving rise to that  $C=-2 \cosh(\alpha)/\alpha $.  Figure~\ref{Gauss} shows the vorticity distribution,  the energy and the angular momentum as functions of $\alpha$. 
Note that the condition $u_\phi(\theta=\pi/2)=0 $ leads to unphysical  results
$u_\phi(\theta=0) =u_\phi(\theta=\pi)=\infty$ for $\alpha \neq 0$.

\begin{figure}[htp]
	\centering
	\includegraphics[width=6cm]{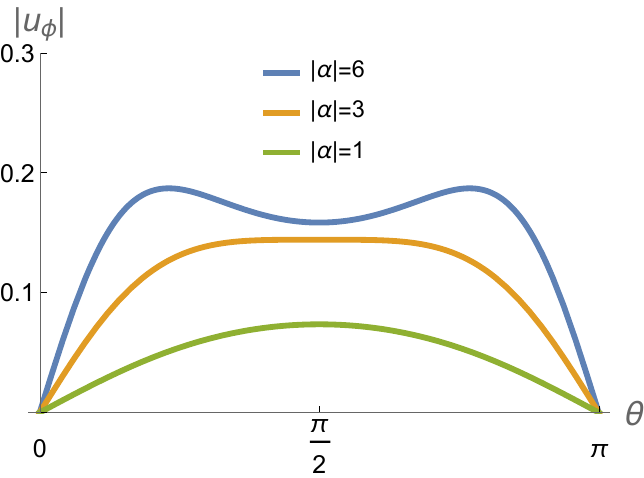}
	\caption{The amplitude of the fluid  velocity field $u_\phi$ for different values of $|\alpha|$. }  
	\label{velocityGaussian}
\end{figure}

Similar to the fluid velocity field for the dipole state,  for the Gaussian state the peak fluid velocity appears on the equator for small values of  $|\alpha|$ and moves towards to poles as $|\alpha|$ exceeds a certain value of $|\alpha_c|$ (Fig.~\ref{velocityGaussian}), where $|\alpha_c|$ satisfies ${\alpha_c}^2  - \cosh (\alpha_c)+1  =0$. The amplitude of the fluid velocity on the equator is $|u_\phi(\theta=\pi/2)| =(1-1/\cosh \alpha)/(2 \pi)$ and reaches the maximum value $1/2\pi$ when $|\alpha| \to \infty$.

\section{Monte Carlo Simulations }

In this section, we show numerical results of Monte Carlo (MC) simulations of a large number of point vortices on a unit sphere.  The simulations are performed for a range of increasing energies and at fixed angular momentum $L_{\zeta}= 0$ for the sandwich and the quadrupole state and  $L_{\zeta}\neq0$ for the dipole state.

\subsection{Microcanonical Monte Carlo}

In our microcanonical Monte Carlo (MC) simulations, we adopt the general scheme developed in  Refs.~\cite{creutz1983,SmithPRL, SmithONeil, clusteringYu}, where the total energy $E$ and the angular momentum $L_{\zeta}$ are fixed within narrow shells. For vortices on a unit sphere,  $q_i=\cos\theta_i$ and $p_i=\phi_i$ are canonical variables~\cite{Kimura99}. In order to keep sampling vortices uniformly in the phase space, at each Monte Carlo step, we randomly  choose a vortex dipole at positions $\bm r^{+}(\theta_i,\phi_i)$ and  $\bm r^{-}(\theta_i,\phi_i)$  and move it to $\bm r^{+}(\theta'_i,\phi'_i)$ and  $\bm r^{-}(\theta'_i,\phi'_i)$,  where $\phi'= p $, $\theta'_i=\arccos(q)$ with $p\in[0,2\pi]$ and $q\in[-1,1]$being independent and  uniformly distributed random variables.  The new configuration is accepted if  the updated energy $ E'$ and the updated angular momentum $L'_{\zeta}$ are within the narrow shells. Otherwise, the vortex  configuration remains the same as the previous one.

%Next, we calculate the energy $E'$ and angular momentum $L'$ of the new configuration. The changes in this quantities are $\Delta E = E'- E$, $\Delta L = L'- L$.
% Get the energy and angular momentum of demo, $E_{\rm d}'=E_{\rm d}- \Delta E$, $L'_{\rm d}=L_{\rm d}- \Delta L$.
% 

%First of all, we prepare an initial  configuration of  $N$ vortices with  given  energy $E$ and angular momentum $L$ on a sphere.  A demo is coupled to this system to simulate the thermodynamic fluctuations of the ensemble. The demo has very little energy $E_{\rm d}$ and angular momentum $L_{\rm d}$, the constraints are $|L_{\rm d}|<L_{\rm m}, |E_{\rm d}|<E_{\rm m}$. The typical scale of demo is $E_{\rm m}=5E_0\times 10^{-3}$, $L_{\rm m}=2N \times 10^{-3}$. %The initial energy and angular momentum of demo are zero.
%To ensure the conservation of total energy and angular momentum, it meets
%\begin{align}
%     E_{\rm Total} &=E+E_{\rm d},\\
%     L_{\rm Total} &=L+L_{\rm d}.
%\end{align}
% In the process of finding the maximum entropy state that satisfies given conditions, we do the Monte Carlo steps (MCS). Each MCS consists of the following procedures.

\subsection{The sandwich state vs. the quadrupole state} 
\begin{figure}[htp]  
	\centering
	\includegraphics[width=8cm]{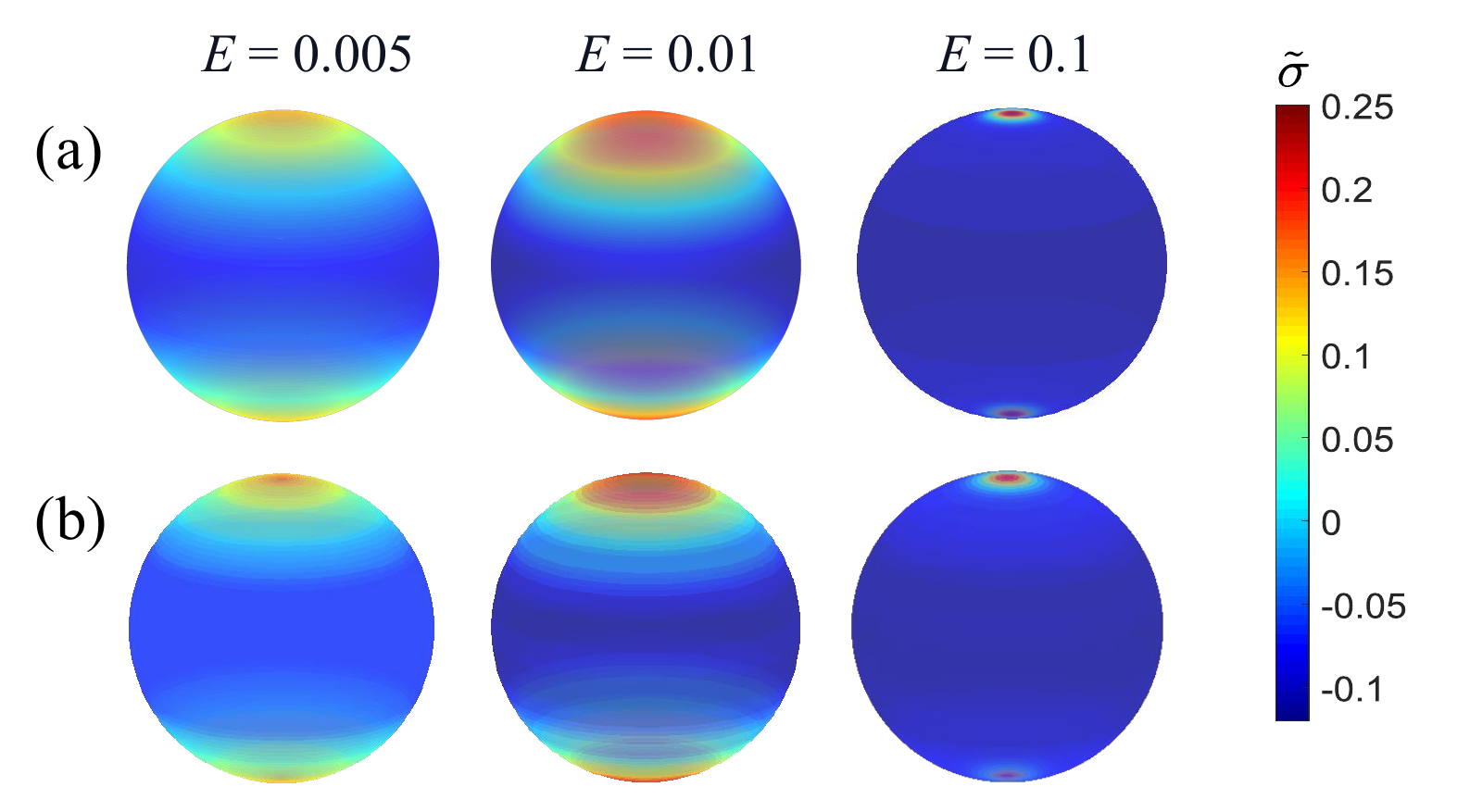} 
	\includegraphics[width=8cm]{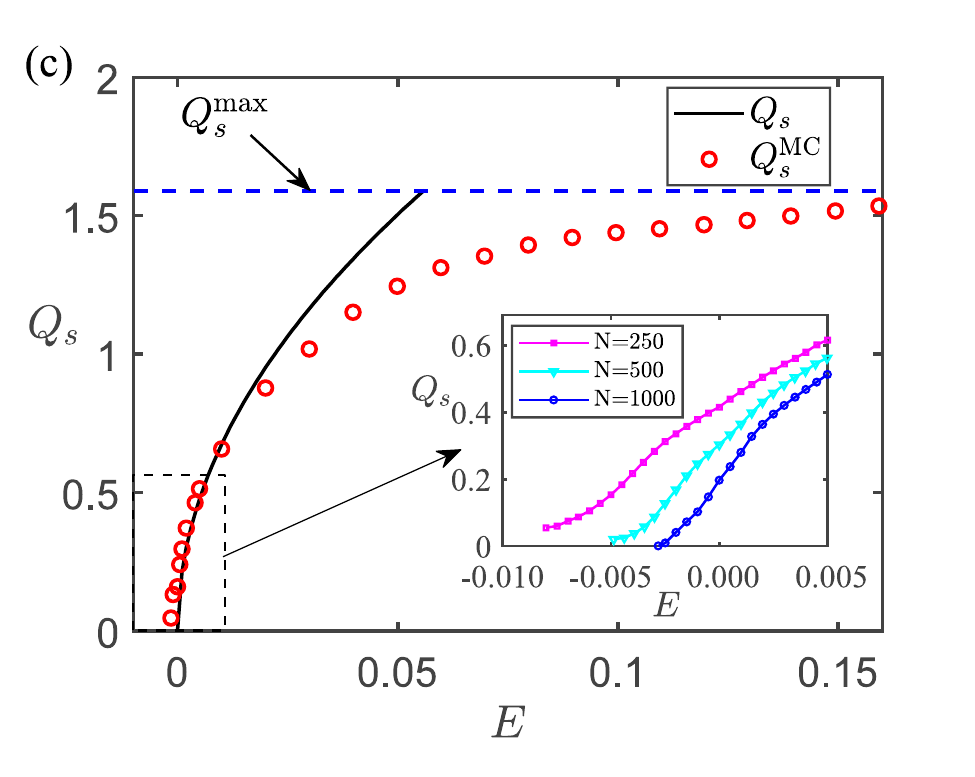}  
	\caption{A comparison between the mean field theory and MC simulations for $N=1000$ point vortices.  (a) and (b) show results from the mean field theory and Monte Carlo sampling for the scaled vorticity $\tilde{\sigma}$, respectively. 
		Here $\widetilde{\sigma} =\sigma$ for $E = 0.005$, $E=0.01$, and $\widetilde {\sigma} =  \sigma/10 $ for $E= 0.1$.  The purpose for introducing the scaled vorticity $\tilde{\sigma}$ is to present the large energy range on a single color map. 0 (c) The quadrupole moment $Q_s$ for increasing energy $E$ (circle red) and  the mean field prediction Eq.~\eqref{quadrupolemoment} (solid blue).  The quadrupole moment is well bounded by the maximum possible value $Q^{\rm max}_s$ predicted by the mean field theory.  The inset in (c) shows the $N$-dependence behavior of $Q_s$ in the vicinity of the clustering transition.  For finite $N$, the transition energy $E_c$ is finite and approaches zero as $N \rightarrow \infty$.  The finite values of $Q_s \sim 1/\sqrt{N}$ below the transition energy are nearly energy independent due to uncorrelated fluctuations. The data presented here are taken in runs of $40000$ MC steps at each value of energy.} 
	\label{Qmsandfig}
\end{figure}

By appropriately choosing initial vortex distributions for given energies and zero angular momentum,  the sandwich state appears at the end of our MC sampling. A comparison between the analytical predictions and the numerical results is shown in Fig.~\ref{Qmsandfig} and excellent agreements are found.  
Caution should be taken here,  it is that the initial configures are chosen to be not far from the sandwich state due to our limited MC sampling efficiency. It is difficult for our current MC sampling method to select the sandwich state but not the quadrupole state from a random initial vortex distribution within the energy and angular momentum constraints.   

\begin{figure}[htp]
	\centering 
	\includegraphics[width=7.99cm]{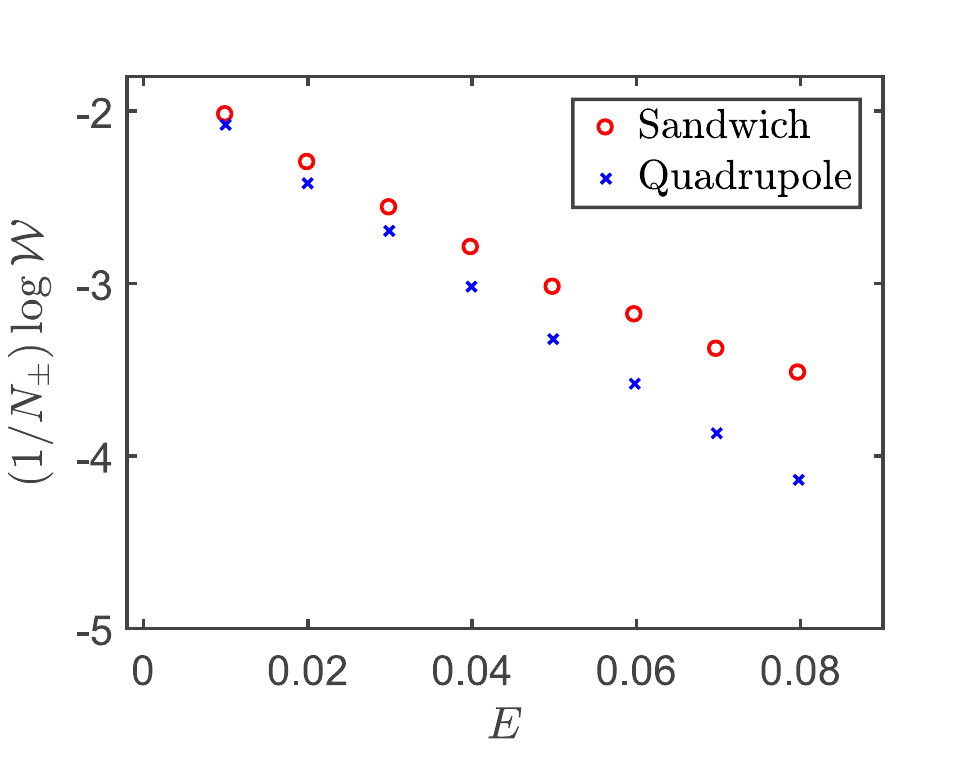} 
	\caption{A comparison between the entropy of the quadrupole state and the sandwich state.  Here $\tilde{N}=32$ and the quadrupole and the sandwich states are prepared via MC sampling for $N=4000$ point vortices with initial vortex distributions very closed to the quadrupole  state and the sandwich state, respectively.   The entropy is negative because of the artificial choice of the minimum area in the phase space, and only the entropy difference between the two states has physical meaning. }
	\label{Qsand} 
\end{figure}

To verify that the sandwich state is the maximum entropy state for large but finite $N$, we numerically evaluate the entropy values of the sandwich state and the quadrupole state at a certain energy range. The entropy values are evaluated using Eq.~\eqref{numberofstates}.  We uniformly divide the phase space of a vortex on a unit sphere  into a large number of cells labeled by $[q_i,p_j]$ with $q_i=-1+2(i-1)/\tilde{N}$, $p_j=2\pi (j-1)/\tilde{N}$ and $j=1,...\tilde{N}+1$,  where  $(i,j)$ denotes the cell index and $\tilde{N}^2$ is number of the cells. The area of a cell is then $a_i=4 \pi/\tilde{N}^2$.  In terms of spherical coordinates, the cell labels are $[\theta_i,\phi_i]$ with $\theta_i=\arccos q_i$ and $\phi_j=p_i$. Note that here $\tilde{N}$ should be properly chosen such that most cells contain a sufficient large number of vortices. 
%The grid size decreases properly  with increasing the energy, so that the entropy is closely related to the choice of grids.  
%As it follows the entropy per vortex is $S=(1/N)\log \mathcal{W}$. 
Using this method we confirm  that the sandwich state has higher entropy than the quadrupole state (Fig.\ref{Qsand}).

%{\color{red}
%
%We denote the number of vortices in a cell of each species as  $N_{i,\hat{\sigma}}$ with $\hat{\sigma}=\pm$ and the number of states is then given by~\cite{Joyce1973, Maestrini2019}
%\begin{align}
%\mathcal{W}\left(\{N_{i,\hat{\sigma}} \} \right)= \prod_{\hat{\sigma}} \left\{ \prod_i \frac{N_{\hat{\sigma}}!}{N_{i,\hat{\sigma}}!}\left(\frac{a_i}{\Omega} \right)^{N_{i,\hat{\sigma}}} \right\} ,
%\label{numberofstates}
%\end{align}
%where $\Omega$ is the area of the unit sphere.  As it follows the entropy per vortex is $S=(1/N)\log \mathcal{W}$. Using this method we confirm  that the sandwich state has higher entropy than the quadrupole state (Fig.\ref{Qsand}).  
%}
 %This can be verified by measuring directly the entropy of the two states for a certain range of energies [Fig.\ref{Qsand}].  
%In order to verify this analytical result numerically, we also obtain the quadrupole state at certain energies by suitable MC sampling.   

% 

\subsection{The dipole state}
\begin{figure}[htp] 
	\centering
	\includegraphics[width=8.1cm]{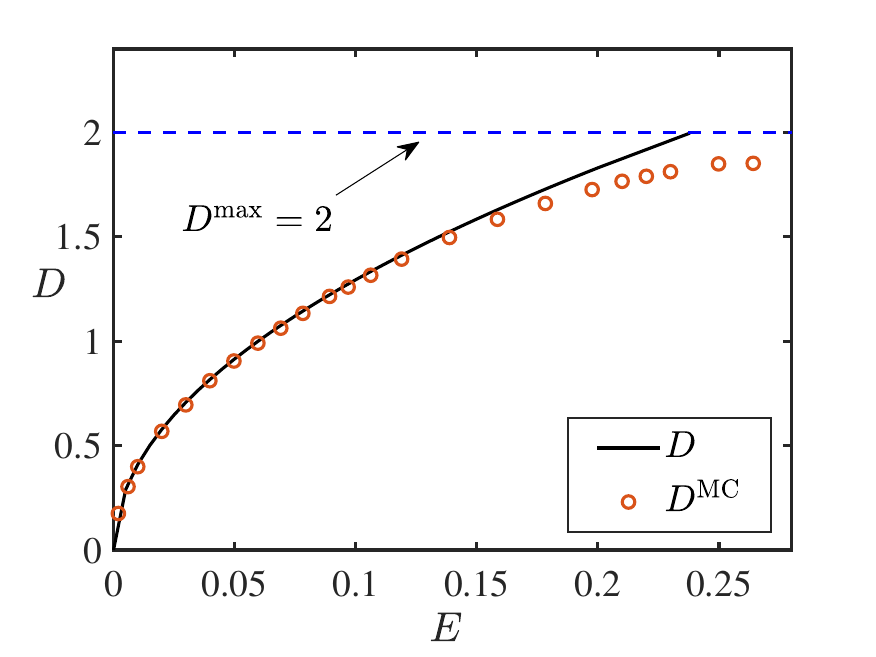}  
	\caption{A comparison between the analytical prediction  $D=|D_{\zeta}|=4\sqrt{\pi/3}\sqrt{E}$ [Eq.~\eqref{dipolemoment} ](solid black) and the MC sampling for $N=1000$ point vortices (circles) in the dipole state.  The dashed blue line marks the upper bound of  the dipole moment $D^{\rm max}=2$ which is reached at the supercondensate point. Note that 
		the angular momentum $L_{\zeta}=4\sqrt{\pi /3}\sqrt{E}$ also increases as increasing energy for this rotation free ($\omega=0$) dipole state.  } 
	\label{LE}
\end{figure}

Our perturbation analysis at low energies (Sec.~\ref{diplestate})  shows that for $L_{\zeta} \neq 0$ and  $\omega=0$ the dipole state maximizes the entropy for a neutral vortex system on a sphere. 
This is different from neutral vortices confined to a disc where the dipole state carries zero angular momentum~\cite{clusteringYu}.  In this subsection,  we present the MC simulations and  compare them with our analytical predictions, showing excellent agreements (Fig.\ref{LE}). 

 \section{Conclusion and discussion}
We investigate rotation free Onsager vortex clustered states of point vortices confined on a sphere.  We find that the sandwich state is the most probable state with zero angular momentum and is characterized by the quadrupole moment tensor.   The dipole state is the maximum entropy state with finite angular momentum and is characterized by nonzero dipole moment. Our mean field analytical predictions agree well with the Monte Carlo simulations. In Ref.~\cite{Guo2021}, it is shown that the point-vortex  dynamics combined removing low energy vortex pairs drives an initial uniform vortex state to the quadruple state. This result is not necessarily inconsistent with our findings, since the quadruple distribution also solves the self-consistent equation and hence is a local maximum entropy state. In the situation described in Ref.~\cite{Guo2021}, vortices may be dynamically trapped in a local equilibrium state. To explore the entropy barrier between the sandwich state and the quadruple state and its effect on the dynamical transitions between the two states deserves future investigations.

 \subsection*{\textbf{Acknowledgement}}
We thank C. Ma and Y. Xiong for useful discussions. X.Y. acknowledges support from the National Natural Science Foundation of China (Grant No. 12175215, Grant No. 12475041), the National Key Research and Development Program of China (Grant No. 2022YFA 1405300) and  NSAF (Grant No. U2330401).

 \appendix 

\renewcommand\thefigure{\thesection} 

%\renewcommand{\theequation}{S\arabic{equation}} 
%\renewcommand{\thefigure}{S\arabic{figure}}
% Maths and other shortcuts

%\setcounter{secnumdepth}{0}
 
% \section{Appendix}

 \section{Comparison of entropy between $\ell=2$ and $\ell>2$ modes}
\label{A}
To leading order in $\delta E$, the entropy for different modes is obtained in the main text  and reads
\bea
S= S_0-\frac{1}{E_0}\left[n_0\beta^2_{c,\ell} \int \d\Omega \, \psi^2_{\ell m} \right] \delta E+\smallO{( \delta E)},
\label{entropychangeleadingorderappendix}
\eea  
where $\beta_{c,\ell}=-\ell(\ell+1)/(2n_0)$.
Let us consider $\ell=2,m=0$ and $\ell=3,m=0$ states:   
\begin{align}
	\psi_{20}=& c_{20} {\rm Y}_{20}= \sqrt{ \frac{E_0}3} \frac{1}{4} \sqrt{\frac 5\pi}(3\cos^2\theta-1), \\
	\psi_{30}=& c_{30} {\rm Y}_{30}=\sqrt{ \frac{E_0}6}  \frac{1}{4}  \sqrt{\frac{7}{\pi }} \left(5 \cos^3\theta-3 \cos\theta \right) ,
\end{align}   
and evaluate the corresponding  values of the entropy  using  Eq.~\eqref{entropychangeleadingorderappendix}.  We find that 
\begin{align}
	 S[{\ell=2,m=0}]= S_0-12\pi \delta E+ \smallO{( \delta E)} ,\\
	S[{\ell=3,m=0}]= S_0-24\pi \delta E + \smallO{( \delta E)} , 
\end{align} 
and  clearly $ S[{\ell=2,m=0}]>S[{\ell=3,m=0}]$.  In general, since $\beta^2_{c,\ell>2}-\beta^2_{c,\ell=2}\sim [(\ell-2)^2+5(\ell-2)]^2$ and $\int \d\Omega \, \psi^2_{\ell m}  \sim  \mathcal{O}{(1)} $, we can conclude that $\ell =2$ have larger entropy than the modes with $\ell >2$.

\section{Boundary conditions}
\label{b}
In the main text, we report that at high energies the vortex distributions near the poles and the equator (in the stereographic coordinates) can be well-approximated by exact solutions to Eq.~\eqref{Poissonz1} [or Eq~\eqref{Poissonz3}] describing distributions of chiral vortices confined to a disc with proper boundary conditions.  Here we show that these boundary conditions are induced be the flow generated by the vortices on the sphere and can be obtained from the global stream function Eq.~\eqref{streamfunctionsandwich} at low energies.      

The relevant stream function  Eq.~\eqref{streamfunctionsandwich} is
\begin{align}
	\psi_{20}=c_{20} {\rm Y}_{20}=\frac{c_{20}}{4} \sqrt{\frac5\pi}(3\cos^2\theta-1),  
	\label{astreamfucntion}
\end{align}
and hence the velocity field reads
\begin{align}
	u_\theta=\frac1{\sin \theta }\frac{\partial \psi}{\partial \phi} , \quad u_\phi=-\frac{\partial \psi}{\partial \theta}  .
\end{align}
It is clear that 
\bea 
u_\phi(\theta=0)=u_\phi(\theta=\pi)=u_\phi(\theta=\pi/2)=0.
\label{boundaryconditonsphere}
\eea
By the stereographic projection [$\rho=\cot(\theta/2)$],  Eq~\eqref{astreamfucntion} becomes 
\begin{align}
	\psi_{20}(\rho)= \frac{c_{20}}{4} \sqrt{\frac5\pi}\left(3 \left(\frac{\rho^2 -1}{\rho^2 +1}\right) ^ 2 -1\right)   
\end{align} 
and the velocity field \eqref{boundaryconditonsphere} induces  that 
\bea
\psi'(\rho)|_{\rho=0}= \psi'(\rho)|_{\rho=1} =0
\eea
which are the adopted boundary conditions in the main text. 

Let us now discuss the irrelevance of the values of $\psi(\rho=0)$ and $\psi(\rho=1)$.   
If applying the boundary condition $\psi'(\rho)|_{\rho=0}=0$ while without specifying the value of $\psi(\rho=0)$, the solution to Eq.~\eqref{poissd1} reads
\begin{align}
	\psi(\rho)=-\frac{2}{\beta} \log  \left(\frac{2 \sqrt{C} }{ \beta  \rho^2 {\rm e}^{\gamma_+} C -2  }\right)
	\label{apsic}
\end{align} 
and the vortex density reads 
\begin{align}
	n_+(\rho)=\exp(-\beta \psi+\gamma_+ )= \frac{4 {\rm e}^{\gamma_+} C}{ (  \beta \rho^2  {\rm e}^{\gamma_+}C -2)^2},
	\label{anc}
\end{align}  
where $C$ is an arbitrary constant.   Clearly here the overall factor ${\rm e}^{\gamma_{+}} C$ can be determined by the normalization condition
\begin{align}
	\frac{1}{2} = \int^{2\pi}_0  \d \phi  \int_0^1  \frac{4\rho}{(1+\rho^2)^2} \, \d \rho \, n_+(\rho), \label{aN1} 
\end{align}  
and the value of $C$ is irrelevant.  In the main text, we choose $\psi(\rho=0)=0$, which gives rise to $C=1$,  for convenience.

Similarly, if applying the boundary condition $\psi'(\rho)|_{\rho=1}=0$ while without specifying the value of $\psi(\rho=1)$, the solution to Eq.~\eqref{Poissiond2} reads
\begin{align} 
	\psi(\rho)=\frac 1\beta \log \left(\frac{  A^2 (4- A^2)\rho^{(-2 + A) }}  {2\beta {\rm e}^{\gamma_-}( A+2- (A -2 ) \rho^{A })^2 }\right)
\end{align} 
and the vortex density is
\bea
n_-(\rho)=\frac{2 A^2  (4 -A^2 ) \rho^{A-2}}{\beta \left(A \rho^A-2 \rho^A+A+2\right)^2},
\label{aned}
\eea  
where $A$ is a constant and  can be determined by the normalization condition 
\begin{align}
	\frac{1}{2} = \int^{2\pi}_0  \d \phi  \int_0^1  \frac{4\rho}{(1+\rho^2)^2} \, \d \rho \, n_-(\rho).
\end{align}  
Hence the value of $\psi(\rho=1)$ is irrelevant and only specifies  the relation between $A$ and $\gamma_{-}$.  With the choice $\psi(\rho=1)=0$ which we made in the main text, we have $A= \sqrt{4 -2 \beta {\rm e}^{\gamma_-}}$ and hence the normalization condition determines $\gamma_-$.  

%\begin{align}
%	\psi(\rho)=\frac 2 \beta \log \left(\frac{2A \rho^{(-1 + A/2 ) }}  {2 + A+ (A -2 ) \rho^{A } }\right) ,\label{apsiedge}
%\end{align}  
%with the boundary condition $ \psi(\rho=1)=\psi'(\rho)|_{\rho=1}=0$.

\section{Normalization dependence of  the onset of  the supercondensation}
\label{c}

\begin{figure}[htp]
	\centering
	\includegraphics[width=4.0cm]{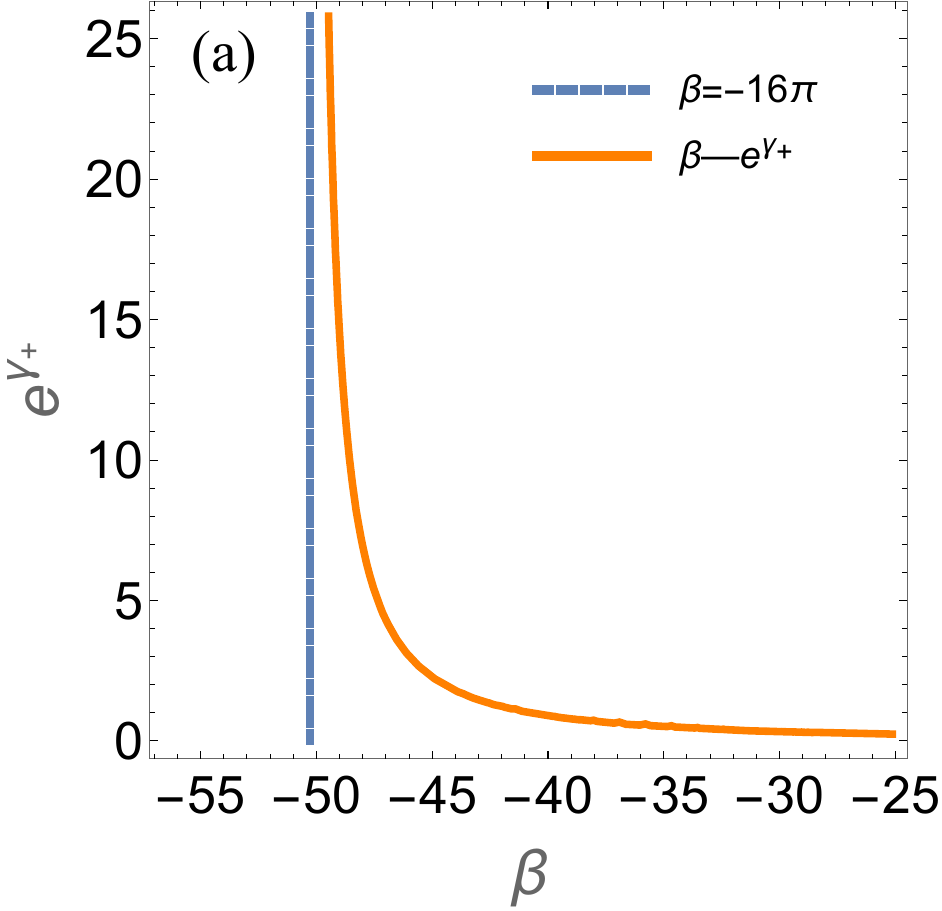}
	\includegraphics[width=4.0cm]{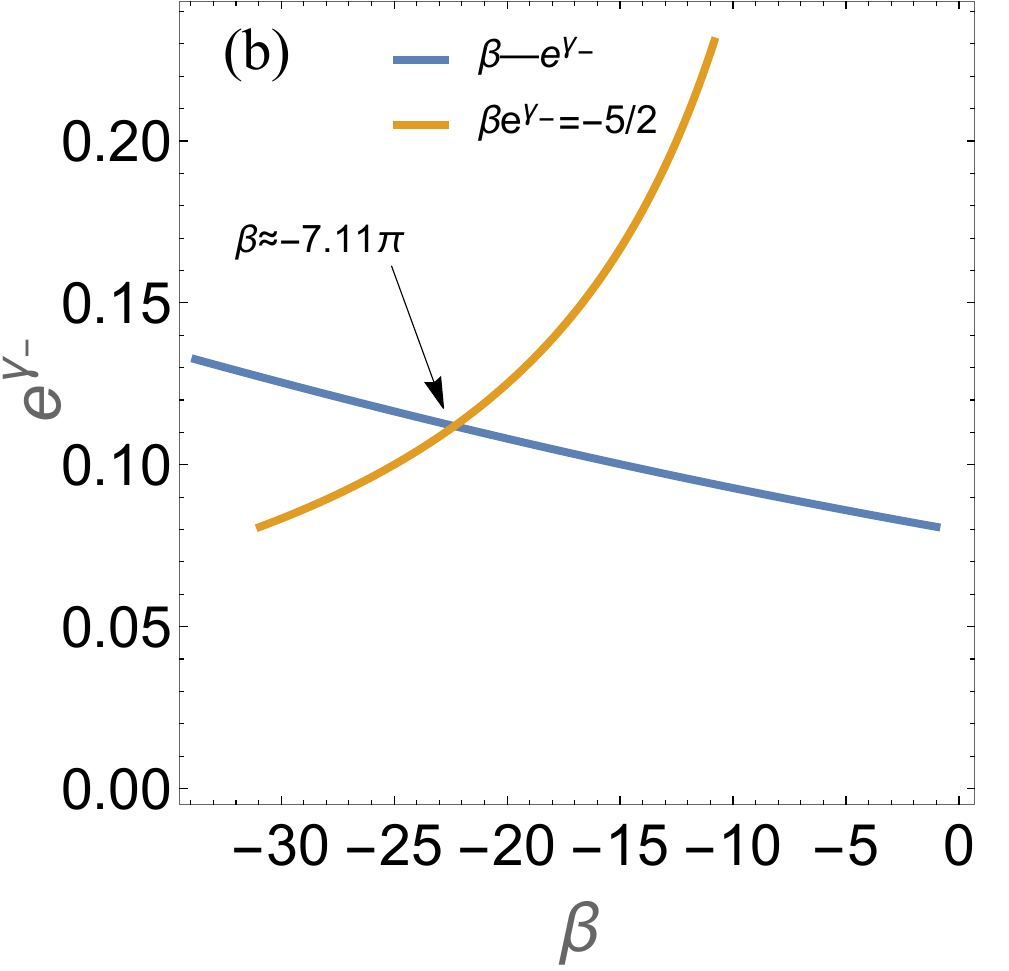}
	\caption{(a) The $\beta-{\rm e}^{\gamma_+}$ curve is determined by normalization Eq.~\eqref{N1} and touches the $\beta=\beta_s=-16\pi$ curve when the supercondensation occurs. (b) The $\beta-{\rm e}^{\gamma_-}$ curve is determined by the normalization Eq.~\eqref{N1} ($n_{+} \rightarrow n_{-}$).  The condition $\beta {\rm e}^{\gamma_-}<-5/2$ ensures $n'_{-}(\rho)|_{\rho=0}=0$. }  
	\label{betag}
\end{figure}

%\label{appendixa}
The value of $\beta$ at which the supercondensation occurs depends on the normalization condition.  Here we show the method we used to determine the value of $\beta_s$ for the sandwich state [see Fig.~\ref{betag}(a)]. 

 \newpage

% \bibliography{Reference}

\end{document}